\documentclass[useAMS,usenatbib]{mn2e}
\usepackage{graphicx}
\usepackage{lscape}
\usepackage{times}
\usepackage{amsmath}
\usepackage{amsfonts}
\usepackage{amssymb}
\usepackage{multirow}
\usepackage{longtable}
\usepackage{afterpage}
\RequirePackage{color}

\title[On the size of giant planets]{On the structure and evolution of planets and their host stars -- 
effects of various heating mechanisms on the size of giant gas planets
}

\author[M. Y\i ld\i z, Z. \c{C}elik Orhan, C. Kayhan and G.E. Turkoglu]{M. Y\i ld\i z$^{1}$\thanks{E-mail:
mutlu.yildiz@ege.edu.tr}, Z. \c{C}elik Orhan$^{1}$, C. Kayhan$^{1}$ and G.E. Turkoglu$^{2}$\\
$^{1}$Department of Astronomy and Space Sciences, Science Faculty, Ege University, 35100 Bornova, \.Izmir, Turkey.\\
$^{2}$ University of Guelph, Department of Human Health and Nutritional Sciences and Department of Physics, N1G 2W1 Guelph, Ontario, Canada.
}

\begin{document}

\date{Accepted 2014 December 15. Received 2013 November 14; in original form 2014 April 11}

\newcommand{\wrt}{with respect to }
\newcommand{\MS}{{\rm ~M}\ifmmode_{\sun}\else$_{\sun}$~\fi}
\newcommand{\RS}{{\rm ~R}\ifmmode_{\sun}\else$_{\sun}$~\fi}
\newcommand{\LS}{{\rm ~L}\ifmmode_{\sun}\else$_{\sun}$~\fi}
\newcommand{\MSbit}{{\rm M}\ifmmode_{\sun}\else$_{\sun}$\fi}
\newcommand{\RSbit}{{\rm R}\ifmmode_{\sun}\else$_{\sun}$\fi}
\newcommand{\LSbit}{{\rm L}\ifmmode_{\sun}\else$_{\sun}$\fi}
\newcommand{\RZ}{{R}\ifmmode_{\rm ZAMS}\else$_{\rm ZAMS}$~\fi}
\newcommand{\RT}{{R}\ifmmode_{\rm TAMS}\else$_{\rm TAMS}$~\fi}

\newcommand{\MJ}{{M}\ifmmode_{\rm J}~\else$_{\rm J}$~\fi}
\newcommand{\RJ}{{R}\ifmmode_{\rm J}~\else$_{\rm J}$~\fi}
\newcommand{\MJbit}{{M}\ifmmode_{\rm J}\else$_{\rm J}$\fi}
\newcommand{\RJbit}{{R}\ifmmode_{\rm J}\else$_{\rm J}$\fi}

\maketitle

\label{firstpage}

\begin{abstract}
{
It is already stated in the previous studies that 
the radius of the giant planets is affected by stellar irradiation.
The confirmed relation between radius and incident flux depends on planetary mass intervals. 
In this 
study, we show that there is a single relation between radius and irradiated energy per gram per second ($l_-$),
for all mass intervals. 
There is an extra increase in radius of planets if $l_-$ is higher than 
1100 times energy received by the Earth ($l_\oplus$).
This is likely  due to dissociation of molecules.
The tidal interaction as a heating mechanism is also considered and found that its maximum effect on 
the inflation of planets is about 15 per cent.
We also compute age and heavy element abundances from the properties of host stars, given in the TEPCat 
catalogue (Southworth 2011). The metallicity given in the literature is as [Fe/H]. However, the most 
abundant element is oxygen, and there is a reverse relation between the observed abundances
[Fe/H] and [O/Fe]. Therefore, we first compute
[O/H] from [Fe/H] by using observed abundances, and then find heavy element abundance from [O/H].
We also develop a new method for age determination.
Using the ages we find,  
we analyse variation of both radius and mass of the planets with respect to time, and estimate 
the initial mass of the planets from the relation we derive for the first time.
According to our results, the highly irradiated gas giants lose 5 per cent of their mass in every 
1 Gyr. 
}
\end{abstract}


\section{Introduction}
The main difference between planets and stars is that stars yield
energy they radiate from their nuclear fuel, but fusion reactions do not
effectively occur in core of the planets.
This difference leads us to consider structures of planets and stars {as if they are completely different.}
However, the fact that both groups obey hydrostatic equilibrium makes these
objects similar in some respects.  
Although direct light from the planets is not (or very little) received,
their orbital parameters, mass ($M_{\rm p}$) and radius ($R_{\rm p}$) are found for many systems.
These parameters allow us to assess the courses influencing hydrostatic equilibrium,
so that eventually radius becomes very large in many cases. The aim of this study is to
consider planet and host star interaction, and  
to assess the basic mechanisms responsible for the excess in radius of the transiting giant gas planets.
The effective mechanisms we consider are 
irradiation, tides, molecular dissociation, cooling and evaporation. 

The main scientific motivation behind the planetary research arises from our interest
if there is any form of life that exists somewhere else in the Universe, other than Earth. 
Life has not been discovered yet in another planetary system, but there are very good candidates 
taking place in the habitable zone. Such a discovery will change our understanding of the Universe forever.
Maybe life is rule rather than exception. 
The details of conditions in such planetary systems depend on the dynamics of planets. 
In this regard, the data of all kinds are very important. 

Thanks to the Kepler (Borucki et al. 2009; Koch et al. 2010) and CoRoT 
(Michel et al. 2006; Auvergneş$ $ et al. 2009) missions, huge amount of data on the basic properties of 
planets are accessible now. In addition to them, there are several ground-based projects for discovery
of the new planetary systems:
HARPS (Mayor et al. 2003), HAT (Bakos et al. 2002), HATnet (Bakos et al. 2004), 
KELT (Pepper et al. 2012; Siverd et al. 2012), OGLE (Udalski 2003), Qatar (Alsubai et al. 2013), 
SuperWASP (Street et al. 2003), TrES (Alonso et al. 2004), WASP (Pollacco et al. 2006), 
WTS (Cappetta et al. 2012), XO (McCullough et al. 2005).

Number of confirmed planets in 1036 host stars is given as 1706 (Akeson et al. 2013; exoplanetarchive.ipac.caltech.edu), at the time 
of writing this paper. 
Some of these systems are multiplanetary. Their number is 442. 
According to the data in TEPCat, the mass of the planets ranges 1.017 M$_\oplus= $ 0.0032 \MJ (KOI-314 c; 
Kipping et al. 2014) to 69.9 \MJ (LHS 6343 b; Johnson et al. 2011).
The upper mass range consists of brown dwarfs. 
{With its mass of 10.52 \MJ (Maxted et al. 2013), WASP-18b is considered to be the most massive planet
below the brown dwarf regime ($M_{\rm p} > 13 \MJ)$}.
The sizes of planets range 0.296 R$_\oplus = $ 0.027 \RJ  (Kepler-37 b; Barclay et al. 2013) to 2.09 \RJ 
(WASP-79 b; Smalley et al. 2012). The lower mass interval for 
the inflated planets 
is about 0.35--0.4 \MJbit.
Their sizes nearly take the upper part of the radius range, 0.8--1.4 \RJbit.


Luminosity of a star is mainly a measure of how hot its nuclear core is. 
Its radius, however, is also a sensitive function of 
how energy is efficiently transferred as much as how much energy is produced. 
Therefore, radius of a star depends on structure of its inner and outer regions.
Metallicity significantly affects structure of these regions.
Therefore, we develop a new simple method to find the age of the host stars 
as a function of metallicity, in addition to mass and radius 
(see Section 3.2).

The only observed parameter that reflects the internal structure of a giant gas planet with known mass 
is its radius. It has already been confirmed in many studies that the radii of planets are related with the incident flux
(see, e.g. Burrows et al. 2000; Demory \& Seager 2011).
This means that the atmosphere and the most outer regions are effectively heated by the energy
released by the host star. The giant gas planets Jupiter and Saturn in our Solar system
have nearly the minimum radii in comparison with their counterparts because  they  are 
old enough to cool down but not
sufficiently heated by the Sun.


The equation of state (EOS) for planets (Fortney \& Nettelmann 2010) is much more complicated than stellar EOS.
For massive main-sequence (MS) stars, the pressure against the gravity is essentially the ideal gas 
pressure.  It is well known that low-mass stars,
are in average cooler and denser than high-mass stars. 
For these stars, non-ideal effects become important 
because ionized particles feel the fields of surrounding charged particles. 
In comparison, planets are much cooler and denser than the low-mass stars. Therefore, planets
EOS is much more complicated due to non-ideal interactions between the particles they consist of. This is one of the major obstacles that limits our better 
understanding of planetary structure and evolution.

The radius of a star for a given mass essentially depends on distribution of number of particles and temperature
throughout its interior. Both of these parameters are functions of chemical composition and time 
(see, e.g.  Baraffe, Chabrier \& Barman 2008; Howe, Burrows \& Verne 2014; Lopez \& Fortney  2014; Howe, Burrows \& Verne 2014). 
The similar situation is also valid for gas giant planets, 
although star and planet interiors are very different. 
Chemical composition changes from star to star, however, it seems that chemical composition plays a very important role 
in planets (see Section 4.5).
This may be another important
obstacle that limits  our understanding of the structure and evolution of planets.

The particles in the outer regions of the giant gas planets are assumed to be in molecular form 
(Bilger, Rimmer \& Helling  2013; Miguel \& Kaltenegger  2014).
However, some irradiated planets have so high equilibrium temperature,
which is the blackbody temperature of a planet heated only by its parent star,
the form of the material may be atomic rather than the molecular. Since gas 
pressure is primarily a function of number of particles, different material forms
imply different planetary sizes (see Section 4.3).

One of the most efficient heating mechanisms responsible from inflated planets is tidal
interaction between planet and its host star (Wu 2005; Jackson,
Greenberg \& Barnes 2008; Liu, Burrows \& Ibgui et al. 2008). 
This interaction cause to convert orbital 
energy to the internal energy of the most outer regions of the planet (see Section 4.2).  

In this paper, we consider which and how mechanisms are influencing
the size of gas giant planets. 
The paper is organized as follows.
{ In Section~2, the basic observational properties of the transiting planets and their host stars are presented. 
Section 3 is devoted to the methods we develop for determination of 
the metallicity and age of the host stars. 
In Section 4, we in detail consider how various mechanisms affect the 
planetary size.
Finally, in Section 5, we draw our conclusions.

\section{Basic properties of planets and their host  stars}
The data are taken from the TEPCat data base
(www.astro.keele.ac.uk/jkt/tepcat/) 
%
for the transiting planetary systems in 2014 January 6, and listed in the online Table A1.
In Fig. 1, the radii of planets are plotted with respect to the planetary mass.
For low-mass range there is a nearly linear relation between mass and radius.
About 1 \MJbit, the relation changes and radius is essentially independent of mass,
in particular for the range 0.4--4.5 \MJbit.
For this range the average radius is about 1.28 \RJbit. The minimum and maximum radii are
0.775 \RJ  of WASP-59 b (H\'ebrard et al. 2013) and 2.09 \RJ of WASP-79 b (Smalley et al. 2012), 
respectively.
Both of these planets have slightly lower mass than Jupiter:
masses of WASP-59 b and  WASP-79 b are 0.863 \MJ and 0.90 \MJbit, respectively.
The radius of Jupiter is in the lower part of the range. 

Eccentricity of some planetary systems in TEPCat catalogue is given as zero, but their 
eccentricities updated by Knutson et al. (2014) are non-zero. In Table A1, the updated 
eccentricities, very important for tidal interaction (see Section 4.2), are listed.

Radius of a polytropic model for planets is given as $R_{\rm p}\propto M_{\rm p}^{1-1/n}$, where $n$ is 
polytropic index.
The best representation of internal structure of intermediate mass planets 
is with $n=1$, which gives radius as independent of $M$ (Burrows \& Liebert 1993). 
This simple case allows us to construct toy models for the giant gas planets.
More realistic models are given and discussed in Guillot \& Gautier (2014).

\begin{figure}
\includegraphics[width=85mm,angle=-90]{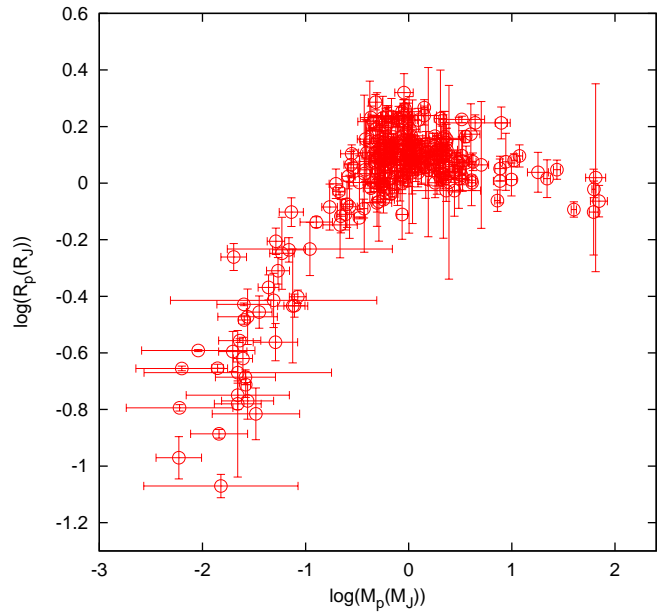}
\caption{ Planetary radius with respect to planetary mass.
}
\end{figure}

We consider the planetary radii in four groups, according to their masses. 
The mass range of the most inflated planets is 0.4--4.5 \MJbit.
Mass of the other two groups are above and below this range. 
This study essentially deals with the planets with $M_{\rm p}> 0.4$ \MJbit.

The effective temperature of the host stars range from 4550 to 7430 K.
The most of them are MS stars, with a limited number of more evolved stars.
Their sizes are in between 0.694 and 6.20 \RSbit.
The mass range of the host stars is 0.75--1.57 \MSbit. 
Age of these planetary systems are found from mass, radius and metallicity of the host stars. 
The method is explained in Section 3.

\section{Estimated heavy element abundance and age for the planetary systems}
\subsection{Estimated heavy element abundance}
\begin{figure}
\includegraphics[width=85mm,angle=-90]{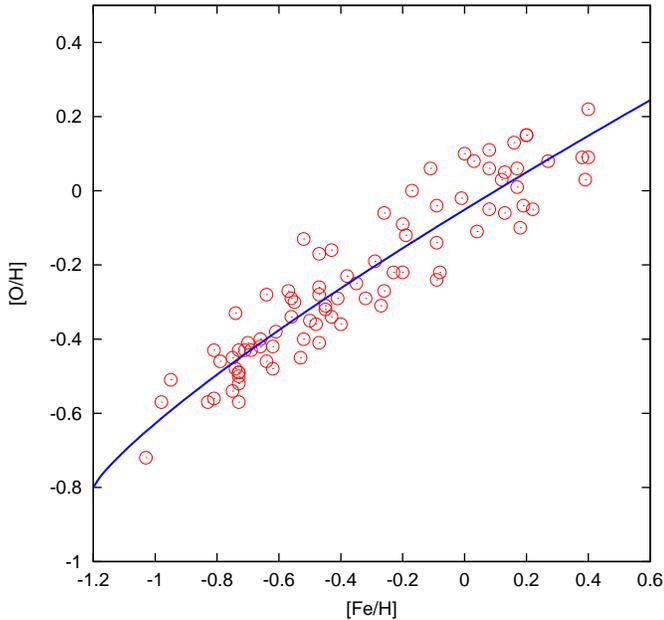}
\caption{ $[{\rm O/H}]$ is plotted with respect to $[{\rm Fe/H}]$. The data are taken 
from Edvardsson et al. (1993). The solid line represents the fitting curve $0.645([{\rm Fe/H}]+1.2)^{0.82}-0.8$. 
}
\end{figure}
\begin{figure}
\includegraphics[width=85mm,angle=-90]{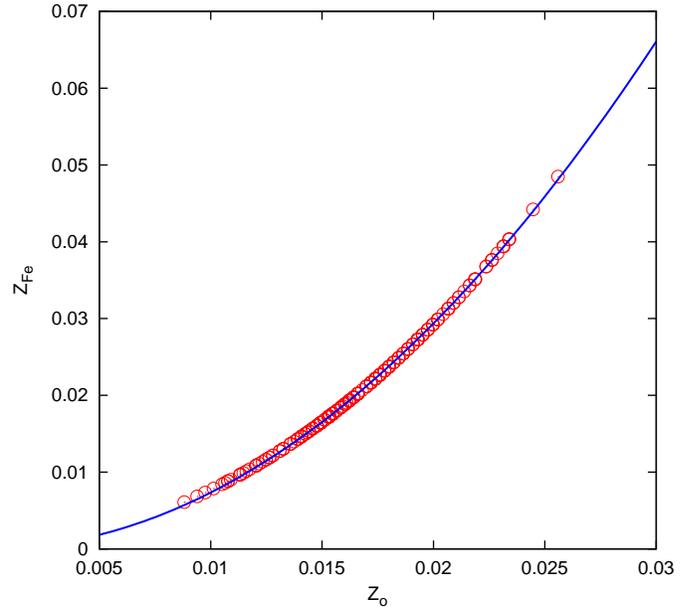}
\caption{ $Z_{\rm Fe}$ is plotted with respect to $Z_{\rm O}$ for the planetary systems. 
The solid line shows the fitting curve $73.4Z_{\rm O}^2$.
}
\end{figure}
\begin{figure}
\includegraphics[width=85mm,angle=-90]{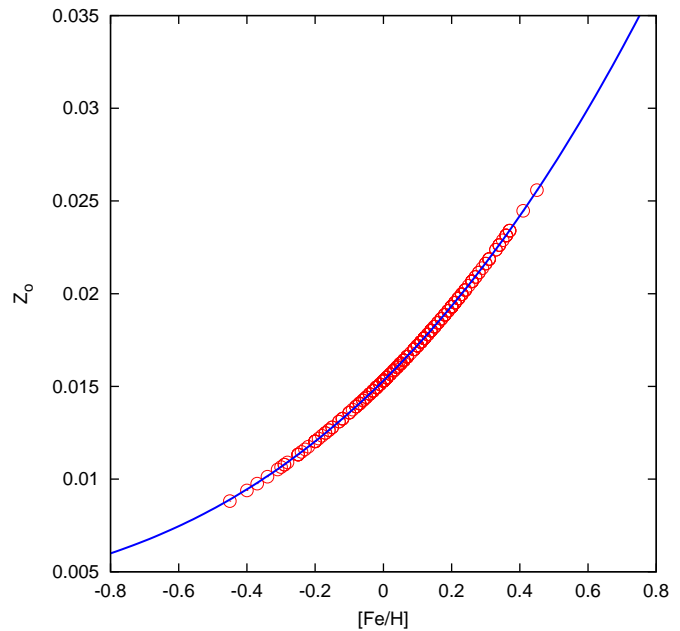}
\caption{ $Z_{\rm O}$ is plotted with respect to $[{\rm Fe/H}]$ for the planetary systems. 
	The solid line is for the derived fitting curve $0.00165([{\rm Fe/H}]+1.916)^3+0.0037$.
}
\end{figure}

MS lifetime and hence age of a star depend on its metallicity ($Z$) and mass ($M$). 
The higher the metallicity is, the greater the age is.
Therefore, 
   observational constraint to $Z$ is needed for a precise age computation. 
$[{\rm Fe/H}]$ abundance is customarily considered as a representative of the total heavy element abundance $Z$ 
   assuming all the heavy element species are enhanced in the same amount relative to the solar abundance. 
   However, oxygen is the most 
   abundant heavy element in normal stars and there is an inverse relation between $[{\rm Fe/H}]$ and 
   $[{\rm O/Fe}]$ abundances, according to the findings of Edvardsson et al. (1993,
 see their fig. 15a-1). They studied solar neighbourhood stars and derived chemical abundances of 189 stars.
They find that [O/Fe] is about 0.2 dex for the metal-poor stars ($[{\rm Fe/H}]\approx -0.2$ dex), and
it is about $-0.2$ dex for the metal-rich stars ($[{\rm Fe/H}]\approx 0.2$ dex).
   Using their data, $[{\rm O/H}]$ is plotted with respect to 
   $[{\rm Fe/H}]$ in Fig. 2. The fitting formula for $[{\rm O/H}]$ is found as
   \begin{equation}
   [{\rm O/H}]=0.645([{\rm Fe/H}]+1.2)^{0.82}-0.8.
   \end{equation}
   Then, we find $Z_{\rm O}$ of the host stars from $[{\rm O/H}]$. We assume a linear relation
between total heavy element and oxygen abundances:
   \begin{equation}
   Z_{\rm O}=10^{[\rm O/H]} {\rm Z_{\sun}}.
   \end{equation}
   In Fig. 3, $Z_{\rm Fe}=10^{[\rm Fe/H]} {\rm Z_{\sun}} $ directly derived from $[{\rm Fe/H}]$ 
is plotted with respect to $Z_{\rm O}$. 
   For low $Z$, the difference between $Z_{\rm Fe}$ and $Z_{\rm O}$ is not very large. For high 
   $Z$, however, the difference is extremely large. While the maximum value of $Z_{\rm Fe}$ is about 0.05, 
the maximum value of $Z_{\rm O}$ is about half of this value,
   0.025. According to the fitting curve $Z_{\rm Fe}\propto Z_{\rm O}^2$, or $Z_{\rm O}\propto Z_{\rm Fe}^{1/2}$.

In Fig. 4, $Z_{\rm O}$ is plotted with respect to [Fe/H]. One can use this figure to find more realistic metallicity 
of any star if its [Fe/H] is observed. 

The metallicity of the planetary systems is between $Z_{\rm O}=0.009$ and $0.026$.
Its mean value is about 0.0169.
Uncertainty in metallicity is mainly due to relation between [O/H] and [Fe/H].  The mean uncertainty in [O/H] is 
about 0.1 dex and this implies a maximum uncertainty in $Z_{\rm O}$ about 25 per cent.
$Z_{\rm O}$ and its uncertainty are listed in Table A1.

   \subsection{Age estimation }
   Radius and luminosity of a star are steadily  increasing
   during the MS evolutionary phase until the totally collapse after the hydrogen fuel is depleted 
near the terminal-age MS (TAMS).
   These are most convenient stellar parameters among non-asteroseismic constraints for
   age determination.
   Only for few  host stars we have asteroseismic constraints and their ages are available in the 
   literature (see Table 1). We have developed a new method for age determination of these stars 
   within the mass range 0.75--1.6 \MSbit, based on stellar mass, radius and metallicity.

   Stellar evolution grids are obtained by using the {\small ANK\.I} stellar evolution code 
   for different composition.
These models are the same models used in Y\i ld\i z et al. (2014a), Y\i ld\i z, 
\c{C}elik Orhan \& Kayhan (2014b) and Y\i ld\i z (2014). 

{  The EOS routines of {\small ANK\.I} take into account Coulomb interaction and solve the Saha equation 
(Y{\i}ld{\i}z \& K{\i}z{\i}lo\u{g}lu 1997).
The radiative opacity is derived from OPAL tables (Iglesias \& Rogers 1996), supplemented by the low temperature tables 
of Ferguson et al. (2005). Nuclear reactions are taken from Angulo et al. (1999) and Caughlan \& Fowler (1988).
The standard mixing length theory is employed for the convection (B\"ohm-Vitense 1958).

The reference models are computed with solar values. From calibration of solar luminosity and 
radius we find that $X_{\sun}=0.7024$, $Z_{\sun}=0.0172$.
The heavy element mixture is taken as 
the solar mixture given by Asplund et al. (2009).
The solar value of the convective parameter $\alpha$ for {\small ANK\.I~} 
is  $1.98$.
}

   According to our stellar evolution understanding, radius ($R$) and luminosity ($L$) of a model
	   are minimum around zero-age-MS (ZAMS) and become maximum near
   TAMS. The TAMS values of $R$ and $L$ 
   are approximately 1.5 and 2 times greater than the ZAMS values, respectively, for the entire mass interval: 
$R_{\rm TAMS}\approx 1.5 R_{\rm ZAMS}$ and   $L_{\rm TAMS}\approx 2 L_{\rm ZAMS} $.

{ We 
consider radius as a function of stellar mass, metallicity and age.
   The difference ($\Delta R$)  between the present radius ($R(t)$) and \RZ is then} a measure of the relative age ($t_{\rm rel}$), 
   defined as
   \begin{equation}
   t_{\rm rel}=\frac{t}{t_{\rm TAMS}},
   \end{equation}
where $t$ is age.
   If $\Delta R=R(t)-R_{\rm ZAMS}$ is about 0.5 \RZ then $t_{\rm rel} \approx 1$ and $t \approx t_{\rm TAMS}$.
   If $\Delta R$ is very small then $t_{\rm rel} \approx 0$ and $t \approx t_{\rm ZAMS}$.

   In order to {find the increase} in radius, \RZ is needed. There is no single relation 
   between mass and radius in the mass range we deal with. For the models with solar composition, 
   one can adopt two different relations for $M_{\rm t}>M_{\rm t\sun}$ and $M_{\rm t}<M_{\rm t\sun}$. $M_{\rm t\sun}$ 
is the transition mass and about $1.3$ \MS for the solar composition.
   We obtain
   \begin{equation}
   \frac{\RZ}{\RS}=0.504\left(M/\MS\right)^{2.02}+0.384
   \end{equation}
   if $M$ is less than $M_{\rm t\sun}$, otherwise,
   \begin{equation}
   \frac{\RZ}{\RS}=1.294\log(M/\MS)^{0.267}.
   \end{equation}
   For arbitrary  $Z$, $M_{\rm t}(Z)=M_{\rm t \sun }(Z/{\rm Z_{\sun}} )^{0.22}$.

   From model computations we confirm that increase in radius is proportional to $t_{\rm rel}^{3/2}$:
   \begin{equation}
   \frac{R(t)}{\RS}=\frac{\RZ}{\RS} +a(M,Z)t_{\rm rel}^{3/2},
   \end{equation}
   where the coefficient $a(M,Z)$ is a function of both $M$ and $Z$.
   It is obtained as 
   \begin{equation}
a(M,Z)=0.114 b\left(\left(\frac{M}{8.8 \rm M_{\sun}}\right)^5+1\right)+(0.222-b)\left(\frac{M}{\rm M_{\sun}}\right)
\end{equation}
where $b$ is a function of $Z$ and found as   
\begin{equation}
b(Z)=\frac{5.297}{3.139+(Z/{\rm Z_{\sun}})^{4.6}}.
\end{equation}
If we are given radius, mass and $Z$, $t_{\rm rel}$ can be found by using equation (6), provided that \RZ is known.
Then, age can be computed from equation (3). We must keep in mind that $t_{\rm TAMS}$ is, beside $M$, also a function 
of $Z$:
\begin{equation}
t=t_{\rm rel}t_{\rm TAMS}(M,Z).
\end{equation}
From comparison of models with different $Z$, we find that
\begin{equation}
t=t_{\rm rel}t_{\rm TAMS}(M,{\rm Z_{\sun}})\left(\frac{Z}{\rm Z_{\sun}}\right)^{0.59}.
\end{equation}
The MS lifetime we derive is as 
\begin{equation}
t_{\rm TAMS}(M,Z=\rm Z_{\sun})=\frac{9.7 \times 10^{9}}{(M/\MS)^{4.14}}\left(0.972+0.0073 \left(\frac{M}{\MS}\right)^{8.4}\right).
\end{equation}

The present method yields the solar age as 4.3 Gyr. This result is in very good agreement with the solar age found by
Bahcall, Pinsonneault \& Wasserburg (1995), 4.57 Gyr.

The ages of the planetary systems for $Z_{\rm O}$ we obtain from the application of the present 
method are given in Table A1. 
The ages range 0.3--11.1 Gyr. The mean value is about 4.2 Gyr.
If we use $Z_{\rm Fe}$ in place of $Z_{\rm O}$, the upper limit for age of the host stars is about 17 Gyr, which 
is much larger than the adopted value of the Galactic age, given as 13.4 $\pm$ 0.8 Gyr by Pasquini et al. (2004).
Uncertainty in age is about 10 per cent.

In the literature, the ages of three planetary systems from asteroseismic inferences are available.
These ages and ages found in this study are listed in Table 1.
The results from two methods are in very good agreement.
\begin{table}
\caption{ Comparison of ages found in this study with the ages from asteroseismic inferences.
}
\centering
\small\addtolength{\tabcolsep}{-3pt}
\begin{tabular}{lcccccrl}
\hline
Star    &$M$&$R$& $Z$&$T_{\rm eff}$ & age& age(seis) &Ref.\\
        & (M$_{\sun}$) &     (R$_{\sun}$)   &    &       (K)& (Gyr)& (Gyr) &\\
\hline
HD 17156 &1.30& 1.49 &0.020 &6079  &  2.7$\pm$0.6 &     2.8 $\pm$ 0.6  & 1 \\

HAT-P-7  &1.51& 1.96 &0.021 &6350  &  1.9$\pm$0.1 &     2.14$\pm$ 0.26 &2 \\
$ $ $ $ "&   &    &      &      &           &        2.21$\pm$ 0.04& 3 \\

Kepler-56&1.32 &4.23 &0.019 &4840  &  3.4$\pm$0.9 &     3.5 $\pm$ 1.3 & 4 \\

\hline
\end{tabular}
$ $ (1) Gilliland et al. (2011), (2) Christensen-Dalsgaard et al. (2010), (3) Oshagh et al. (2013), 
(4) Huber et al. (2013).
\end{table}

\section{The mechanisms influencing planetary radius}
\subsection{Effect of incident flux}
\begin{figure}
\includegraphics[width=85mm,angle=-90]{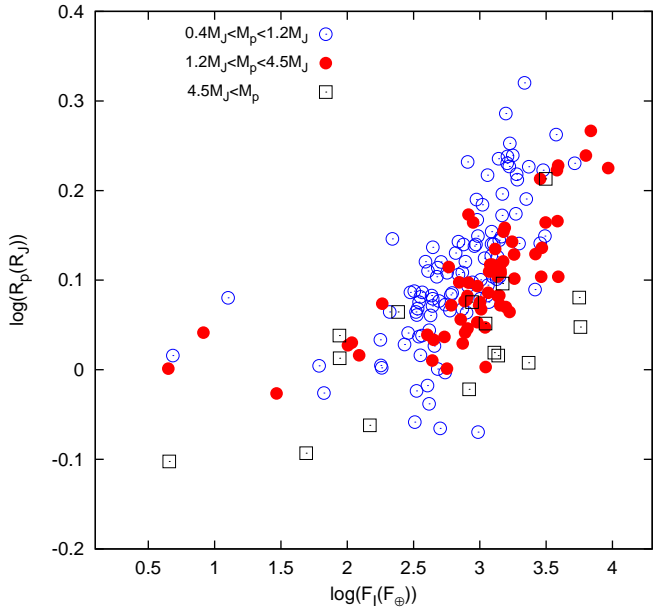}
\caption{ Planetary radius with respect to incident flux in units of incident flux at Earth for different
planetary mass groups. While the circles are for the planets
$0.4$ \MJ $<M_{\rm p} < 1.2$ \MJ and the filled circles show $1.2$ \MJ
$< M_{\rm p} < 4.5$ \MJbit, {the squares are } for $M_{\rm p} > 4.5$ \MJbit.
}
\end{figure}
\begin{figure}
\includegraphics[width=85mm,angle=-90]{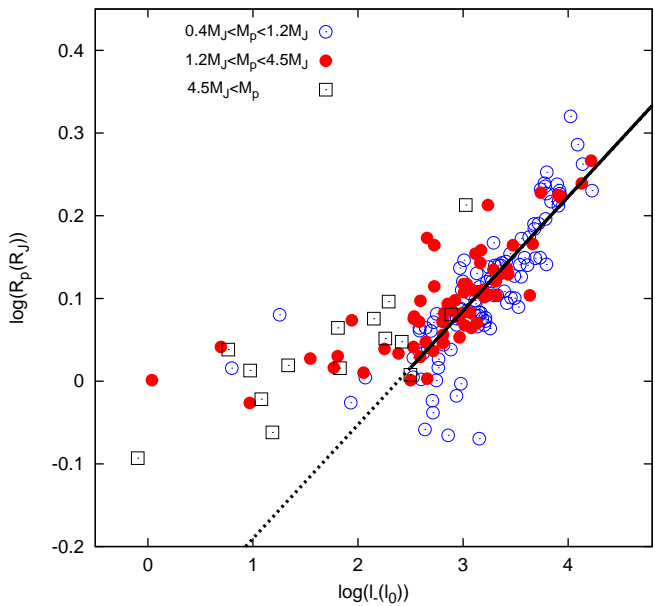}
\caption{ Planetary radius with respect to total energy received by the planet per gram per second.
The fitting line is as ${(0.138\pm0.008)\log(l_-/l_{0})-(0.327\pm0.025)}$. 
The effect of $l_-$ on the radius can be given as $\Delta \log(R_l) = 0.138(\Delta \log(l_-/l_{0})-2.5)$
}
\end{figure}
\begin{figure}
\includegraphics[width=85mm,angle=-90]{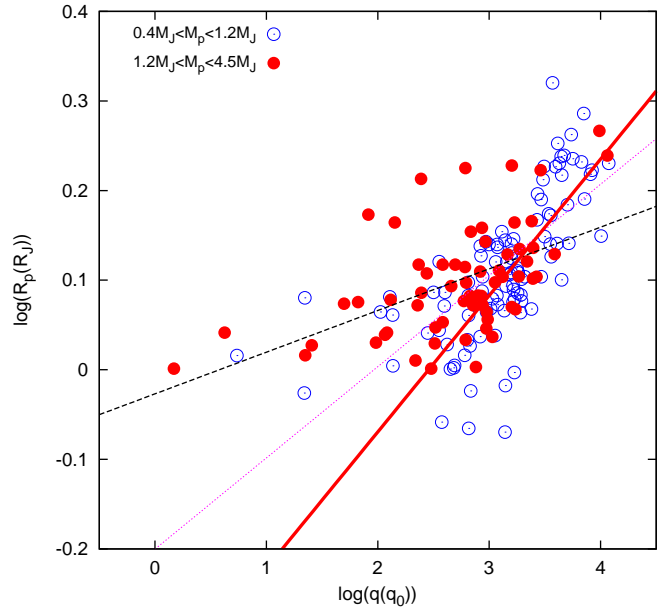}
\caption{ \textcolor{red}{}Planetary radius with respect to heat ($q=l_- t$) in units of $q_0=l_0t_{\sun}$.
The solid line is the fitting line for 0.4--1.2 \MJ and $\log(q/q_0)>2.4$, 
${(0.152\pm 0.014)\log(q/q_0)-(0.374\pm0.048)}$.  
The dotted line is the fitting line for 1.2--4.5 \MJ and $\log(q/q_0)>2.4$,
${(0.102\pm 0.020)\log(q/q_0)-(0.200\pm 0.061)}$. 
The dashed line is also for  1.2--4.5 \MJ but for full range of $\log(q)$,
${0.047\pm 0.010) \log(q/q_0)-(0.027\pm 0.027)}$.        
}
\end{figure}

The effect of incident flux ($F_{\rm I}$) on the size of the planets is widely considered in many studies in the literature
(see, e.g. Guillot et al. 1996;
Burrows et al. 2000; Sudarsky, Burrows \& Hubeny 2003;
Fortney et al. 2007; Demory \& Seager 2011; Weiss et al. 2013). 
The flux heats the outer regions of planets. 
This causes an  increase of temperature of those regions.
If the matter in the heated regions is in the gas form, then pressure increases, and therefore 
the gas planets expand in order to reach hydrostatic equilibrium.
{An alternative explanation} might be that heating prevents cooling of the outer regions of giant gas planets, and therefore 
pressure and eventually their radii remain high.

In Fig. 5, radii of the giant planets of three mass groups are plotted with respect to 
the incident flux in units of flux received by Earth.  
The largest planets are the planets with the highest irradiated energy. 
We consider intermediate mass planets in two subgroups; $0.4$ \MJ  
$<M_{\rm p} < 1.2$ \MJ and $1.2$ \MJ  
$< M_{\rm p} < 4.5$ \MJbit.
The excess in radius is different for different mass groups. For a given incident flux,
the intermediate mass planets have larger radius than the high-mass planets. 
This seems very reasonable because the temperature in the heated region is determined by 
energy per mass, rather than the incident flux. 
Excess in radius must be due to the increase in the pressure of the heated regions. 
The pressure is related to the total
received energy per gram per second.
The total energy per second received by the planet  $L_-$ is
\begin{equation}
L_-= {\rm \pi} R_{\rm p}^2 F_{\rm I}.
\end{equation}
Increase in temperature depends on how much energy is received per gram per second, i.e. on $L_-/M_{\rm p}$ ratio,
assuming proportionality between mass of the heated region and the total planetary mass.
In Fig. 6, $\log(R_{\rm p})$ is plotted with respect to logarithm of $l_-= L_-/M_{\rm p}$ in units of $l_0$,
where $l_0=1.106\times 10^{-4}$ erg g$^{-1}$ s$^{-1}$ is received energy per unit mass and time by a planet with 1 \MJ and 1 \RJ at 1 au in our Solar system.
$l_0$ is 0.379 times the received energy per unit mass and time by Earth ($l_\oplus$).
The inflated planets are the planets heated about 
110--190 ($\log(l_-/l_0)=$2.5--2.7) times more than Earth.
The intermediate and high-mass planets obey the same relation in $\log(R_{\rm p})$ and 
$\log(l_-)$ diagram. Therefore, the relation between $R_{\rm p}$ and $l_-$ is much more explicit than the
relation between $R_{\rm p}$ and $F_{\rm I}$ given in Fig. 5.

As stated above, Fig. 5 shows that the excess in radii of planets receiving the same flux depends on the 
planetary mass. The mean radius difference 
between the planets with 0.4--1.2 \MJ and 1.2--4.5 \MJ is {about $\Delta \log(R_{\rm p}/{\rm R}_{\rm j})\approx 0.1$, for} a given $F_{\rm I}$. 
This means the difference is about 26 per cent. In Fig. 6, however, we note that the mean radii of the planets with 
0.4--1.2 \MJ and 1.2--4.5 \MJ are the same for a given $l_-$.


However, $l_-$ is also related with the ratio of $F_{\rm I}$ to gravity at the surface of the planets:
\begin{equation}
l_-= \frac{{\rm \pi} F_{\rm I}}{M_{\rm p}/R_{\rm p}^2}\propto \frac{F_{\rm I}}{g_{\rm p}}.
\end{equation}
For a given value of  $F_{\rm I}$, increase in radius of a planet also depends on gravitational acceleration
in the expanding outer regions. The weaker the gravity is, the greater the expansion is.
Perhaps, therefore the $R_{\rm p}$--$l_-$ relation is much more definite than the $R_{\rm p}$--$F_{\rm I}$ relation.
We note that the horizontal axis of Fig. 6 also contains $R_{\rm p}$ implicitly. 
If we want to consistently write an expression for radius, $F_{\rm I}/M_{\rm p}$ is the alternative 
of $F_{\rm I}R_{\rm p}^2/M_{\rm p}$.
For this purpose, we derive 
\begin{equation}
\log(\frac{R_{\rm p}}{\RJ})={({0.077\pm0.007}) \log
(\frac {F_{\rm I}/{\rm F}_{\oplus}} {M_{\rm p}/M_{\rm J}})-({0.422\pm0.048})}
\end{equation}
{for $\log(F_{\rm I})> 2.5 \log({\rm F}_{\oplus})$}.

We also test if the total heat added ($q=l_-t$) to the planets during their lifetime 
has any influence on the planetary size.
In Fig. 7, $R_{\rm p}$ is plotted with respect to $q$. The relation between $R_{\rm p}$ and $q$
is not strong as the relation between $R_{\rm p}$ and $l_-$. However, there is a small slope in the 
low-heat region ($q < 3$).

\subsection{Effect of tides}
\begin{figure}
\includegraphics[width=85mm,angle=-90]{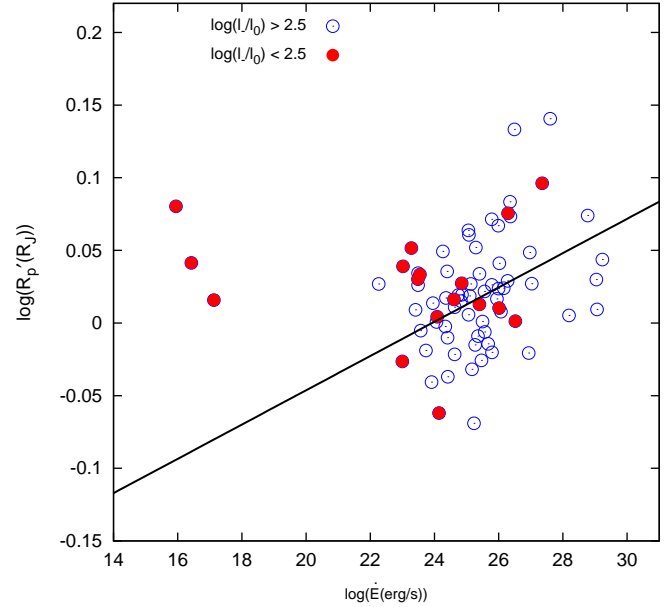}  
\caption{ Planetary radius $R_{\rm p}^\prime$ with respect to energy rate due to tidal interaction between host star and planet
in eccentric orbit. $R_{\rm p}^\prime$ is $R_{\rm p}-\Delta R_l$, where $\Delta R_l$ is the amount of expansion due to 
$l_-$  (see Fig. 6).
The circle and filled circle are for the planets with $\log(l_-/l_{0})>2.5$ and $\log(l_-/l_{0})<2.5$,
respectively.
The fitted line is $(0.012\pm 0.004)\log(\dot{E}/$(erg s$^{-1}))-(0.290\pm 0.100)$ obtained for the range $\log(\dot{E}/$(erg s$^{-1}))>24$.
}
\end{figure}
The tidal interaction between a planet and its host star may heat the planet, if the orbit
is eccentric or the planet rotates asynchronously (Bodenheimer, Lin \& Mardling 2001; Jackson,
Greenberg \& Barnes  2008). 
Since we do not have any information 
about how the planets rotate, we only consider the planetary systems with eccentric orbits. 
The heating mechanism in such systems is studied in, e.g.
Storch \& Lai (2014) and Leconte et al. (2010).
We apply the heating mechanism given by  Storch \& Lai (2014). The rate of energy ($\dot{E}$) converted from
orbital energy to the heat is 
\begin{equation}
\dot{E}=\frac{{\rm d}E_{\rm tide}}{{\rm d}t}= 2K_{\rm p}[N_{\rm a}(e)-\frac{N^2(e)}{\Omega(e)}]
\end{equation}
where $N(e)$, $N_{\rm a}(e)$ and $\Omega(e)$ are only function of the eccentricity ($e$). They are given as
\begin{equation}
N(e)=\frac{1+15/2e^2+45/8e^4+5/16e^6}{(1-e^2)^6},
\end{equation}
\begin{equation}
N_{\rm a}(e)=\frac{1+31/2e^2+255/8e^4+185/16e^6+25/64e^8}{(1-e^2)^{15/2}}
\end{equation}
and
\begin{equation}
\Omega(e)= \frac{1+3e^2+3/8e^4}{(1-e^2)^{9/2}}.
\end{equation} 
Expression for the other parameter $K_{\rm p}$ shown in equation (15) is as
\begin{equation}
K_{\rm p}=\frac{3}{2}k_{\rm 2,p}\Delta t_{\rm p}(\frac{GM^2_{\rm p}}{R_{\rm p}})(\frac{M_{\rm \ast}}{M_{\rm p}})^2(\frac{R_{\rm p}}{a})^6n^2
\end{equation}
where $k_{\rm {2,p}}$, $\Delta t_{\rm p}$ and $n$ are the potential Love number of degree 2, constant time-lag for planet and the 
orbital mean motion, which is $2{\rm \pi}/P_{\rm orb}$, respectively. $k_{\rm {2,p}}\Delta t_{\rm p}$  is taken 
as $2\times 10^{-2}$--$2\times 10^{-3}$ values for giant gas planets. These equations are used for calculating energy of 
planet's tidal effect (Leconte et al. 2010; Storch \& Lai 2014) by adopting the upper 
value of $k_{\rm {2,p}}\Delta t_{\rm p}$. 

The radius of the planets with eccentric orbits and $\log(l_-/l_{0})<2.5$ are plotted with respect to tidal energy rate in Fig. 8. 
In order to see net result of tides, we also plot planets with  $\log(l_-/l_{0})>2.5$ by subtracting the excess in radius due to irradiation
from the observed radius.
The values of 
$\dot{E}$ range from $10^{16}$ 
to $10^{29}$ erg s$^{-1}$. The tidal interaction influences radius of a planet 
if $\dot{E}$ is greater than $10^{22}$ erg s$^{-1}$. The solid line is the fitted line if we assume a linear relation between 
$\log(R_{\rm p}^\prime)$  and $\log(\dot{E})$. 
In order to make the effect of tides clearer, the planets with $\log(l_-/l_{0})>2.5$ 
and $\log(l_-/l_{0}) < 2.5 $ are marked with different symbols. Since irradiation energy is effective on the 
planet radius if $\log(l_-/l_{0}) > 2.5 $, the most inflated planets are the planets with the highest
irradiated energy. The pure tidal effect might be on the planetary systems with 
$\log(l_-/l_{0}) < 2.5 $. The maximum effect of tides seems to be about 10 per cent. 
The same is true for $\log(l_-/l_{0})>2.5$, but the maximum effect is about 15 per cent 
for CoRoT-2 b (Alonso et al. 2008) and WASP-14 b (Joshi et al. 2009).

\subsubsection{Tidal effect and Roche lobe filling factor}
The gravitational acceleration applied by the star on the inner surface of the planet 
($g_{\rm \star}=GM_{\rm \star}/(a-R_{\rm p})^2$) is non-negligible in comparison with the self gravitational acceleration 
($g_{\rm p}=GM_{\rm p}/R_{\rm p}^2$) plus centrifugal acceleration $g_{\rm c}=v^2_{\rm p}/(a-R_{\rm p})$, where
$v_{\rm p}$ is the orbital speed.
When the planet is at periastron,
its nearest part is pulled towards the star, and therefore it gains potential energy. This potential energy is 
converted into kinetic energy, and later into heat when the planet travels towards the apastron. 
However, asynchronous rotation of planet in a circular orbit may also cause to heat the outer regions of planets.

{The} centrifugal acceleration at periastron is 
$g_{\rm cperi}=\omega^2_{\rm peri}a(1-e)$, where $\omega_{\rm peri}=2{\rm \pi}/P\sqrt{(1+e)/(1-e)}$.
$g_{\rm \star peri}=GM_{\rm \star}/a^2(1-e)^2$ 
represents gravitational acceleration due to
the host star at periastron.
The effect of tidal interaction depends on the ratio
of accelerations:
\begin{equation}
r_{\rm peri}=\frac{g_{\rm \star peri}}{g_{\rm p}+g_{\rm cperi}}.
\end{equation}

\begin{figure}
\includegraphics[width=85mm,angle=-90]{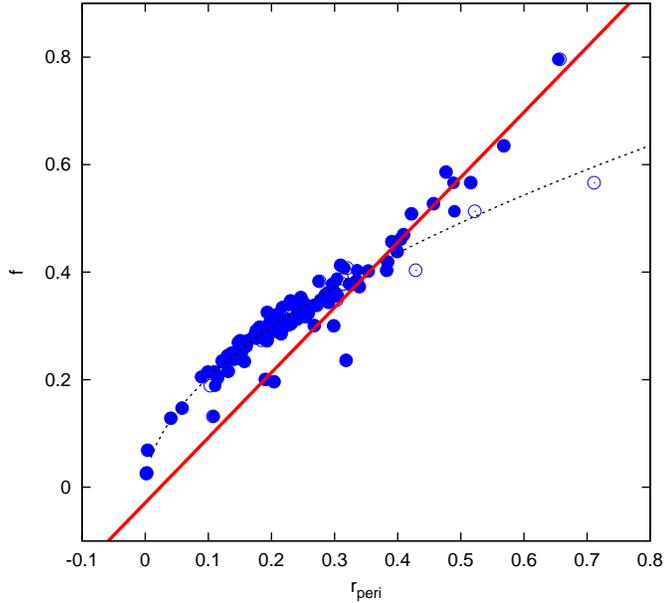}
\caption{ The Roche lobe filling factor, $f$, is plotted with respect to ratio of accelerations, $r_{\rm peri}$
(equation 20) for the gas giant planets with $0.4$ \MJ $<M_{\rm p} < 1.2$ \MJbit.
There are two different relations between $f$ and $r_{\rm peri}$. The curve with dashed line is for the range $f<0.4$, and
the {solid} line is for $f>0.4$. For some of the planetary systems, eccentricities are updated by Knutson et al. (2014).
These systems are represented by circles. We note that two planets with the recent eccentricity 
{values lie} near to the dashed line. 
}
\end{figure}
\begin{figure}
\includegraphics[width=85mm,angle=-90]{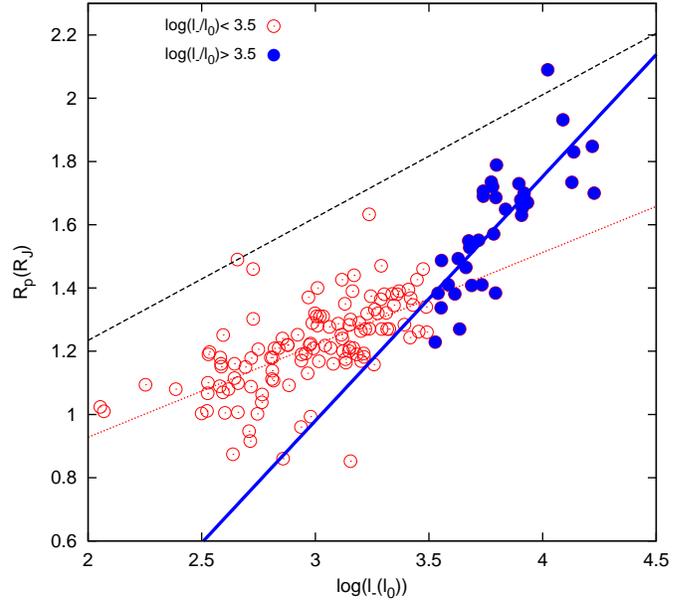}
\caption{ Planetary radius with respect to $\log(l_-)$. The dotted and solid lines are for the ranges
$\log(l_-/l_{0}) < 3.5$ and $\log(l_-/l_{0}) > 3.5$, respectively.
If we consider molecular dissociation, radius of a planet is multiplied by about {33} per cent.
When we multiply line equation for $\log(l_-/l_{0}) < 3.5$ (dotted line) by {1.33} we obtain 
the dashed line (see Section 4.3).
}
\end{figure}
To confirm the effect of tidal interaction on the planetary size, the filling factor ($f$), the ratio
of the planetary radius to the Roche lobe radius, is plotted with respect to $r_{\rm peri}$ in Fig. 9.
The Roche lobe radius is computed by using the expression in Eggleton (1983).
There are two different relations between 
$f$ and $r_{\rm peri}$. The transition occurs at about 
$f=0.35$--$0.40$. This point corresponds $\log(l_-/l_{0}) \approx 3.5$. 
{The curve represents the fit for $f<0.4$ and it is only an extrapolation for $f>0.4$.
For the range $f>0.4$, 
the relation between $f$ and $r_{\rm peri}$ is linear.}

\subsection{Effect of molecular dissociation}
{In Fig. 10, the radii of giant gas planets are plotted with respect to $\log(l_-)$. Two different $\log(l_-)$ ranges
are represented with different symbols. 
There is a good linear relation between planetary radius and $\log(l_-)$ for $\log(l_-/l_{0})<3.5$, with a slope of 0.29. 
The relation 
changes if $\log(l_-/l_{0})>3.5$. There is a relatively sharp slope for the range $\log(l_-/l_{0})>3.5$, about 0.77. 
The reason {for } the change in the slope may be related with the structure of the heated part of the planets.
We note that $T_{\rm eq}>1500$ K, when $\log(l_-/l_{0})>3.5$ and $T_{\rm eq}$ reaches 2700 K for some planets.
$T_{\rm eq}$ values are taken from TEPCat and given in Table A1.
We can expect molecular dissociation at such high temperatures. Maybe, this is the reason for the presence of  two 
very different slopes in Fig. 10 (see below). 
}

%


%



The most abundant species in the surface of the giant gas planets in our Solar system, 
namely Jupiter and Saturn, is H$_2$ molecules (Atreya et al. 2003). 
If H$_2$ molecules in some very hot gas planets dissociate due to various heating mechanisms, 
then we can expect that radii of these planets must be greater than their counterparts.
We can assess the basic effect of molecular dissociation in a simple manner.
Let us consider a gas made of diatomic molecules with $T$, close to but less than
dissociation temperature. 
{
If the gas is heated,
it will expand adiabatically and adopt a new equilibrium state so that pressure at 
a mass element remains constant to good approximation.
The gas transforms from molecular to atomic form in the outer regions and the number density of particles ($n_{\rm f}$) is twice 
the initial value ($n_{\rm i}$). Then, its volume ($V$) increases. 
The relation between $\Delta R=R_{\rm f}-R_{\rm i}$ and $\Delta V=V_{\rm f}-V_{\rm i}$ for a spherical object can be written as
\begin{equation}
\frac{\Delta R}{R_{\rm i}}=\frac{1}{3}\frac{\Delta V}{V_{\rm i}}
\end{equation}
If $n_{\rm f}\approx 2 n_{\rm i}$, then $V_{\rm f}\approx 2 V_{\rm i}$, assuming the ideal gas law is nearly satisfied in 
the expanding outer regions. This implies that $\Delta V= V_{\rm i}$.
Then, from equation (21),
\begin{equation}
R_{\rm f}=R_{\rm i}+\Delta R=1.33 R_{\rm i}.
\end{equation}
%
This implies that $\Delta R$, which is the increase in radius due to increase in number of
particles as a result of molecular dissociation, is $0.33 R$.

If we multiply the fitted line (dotted line) in Fig. 10 by 1.33 (dashed line), then we nearly obtain the maximum effect of 
molecular dissociation on the planetary radius. We note that it limits almost 
all the data (except WASP-79 b) and there are some planets very close to this line. 
}
\subsection{Variation of planetary radius in time -- effect of cooling }
Interior of protoplanets are very hot. Their internal temperature decreases 
in time as a result of the cooling process, at a rate depending on planetary mass and irradiation energy. 
In this section, we consider if the effect of cooling is seen in the current data and 
try to determine  if cooling rate also depends on 
the irradiated flux. To do these, we divide the planets within the mass range
$0.4$ \MJ $<M_{\rm p} < 4.5$ \MJ
into four subgroups according to their $l_-$.
The properties of these subgroups are given in Table 2.
$\ln (R_{\rm p})$ is plotted with respect to age in Fig. 11 for these subgroups.
Also shown in this figure are the fitted lines. The slopes of these lines are different for different $l_-$ intervals
and decrease as $l_-$ reduces.

The alternative expression to the time variation of radius is that planetary radius shows extra dependence 
on stellar effective temperature, for example.  
High effective temperature of host stars means high energetic photons throughout irradiation energy.
Maybe, heating also depends on energy content of photons as well as total irradiation flux.
However, we see a similar time variation also for planetary mass (see Fig. 12). This leads us 
to consider that the planetary {radii most probably change} due to heating and cooling mechanisms (see below).

\begin{figure}
\includegraphics[width=85mm,angle=-90]{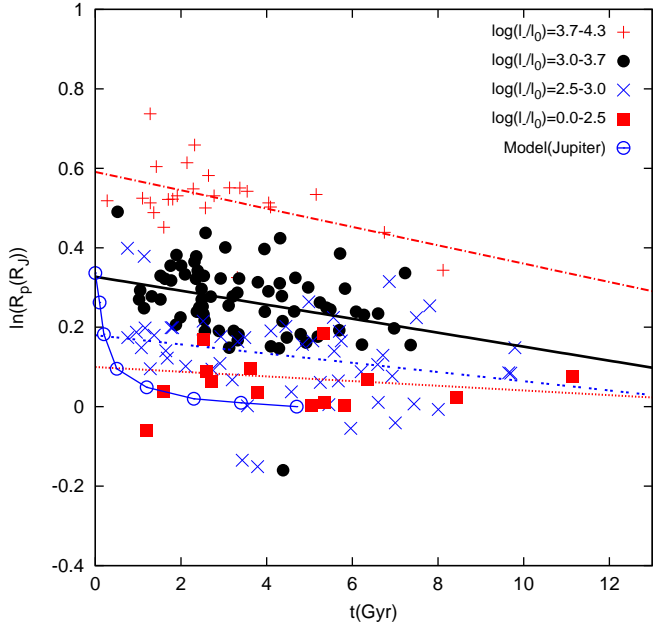}
\caption{ Planetary radius with respect to age for different $\log(l_-)$ intervals.
The greater the $\log(l_-)$ value is, the sharper the slope is. The curve with circle 
is the cooling curve for a Jupiter model of Nettelmann et al. (2012). If there is no remarkable 
irradiation, the cooling is relatively fast and lasts about 1--2 Gyr. 
We notice that
the present slope of the curve is in good agreement with the slope for the planets with the 
lowest irradiation energy.
}
\end{figure}
\begin{table}
\caption{ Time variation of mean radius for different $l_-$ intervals.
The mean radii and ages are given in the second and third columns, respectively. 
The slopes of fitted lines in Fig. 11 are listed in column four. 
The fifth column is for the constants of these lines.
}
\centering
\small\addtolength{\tabcolsep}{-3pt}
\begin{tabular}{rcccc}
\hline
$l_-$ & $\overline{R}(\RJbit)$ & $\overline{t}$(Gyr) &$ \frac{{\rm d} \ln R}{{\rm d}  t_9} $ &  $\ln(R_0/\RJbit)$  \\  
\hline

1.0-2.5  &    1.07        &    5.23       & {-0.001$\pm$0.007} & {0.052$\pm$0.037}      \\
2.5-3.0  &    1.19        &    5.49       & {-0.012$\pm$0.006 }& {0.180$\pm$0.030}      \\
3.0-3.7  &    1.28        &    4.33       & {-0.018$\pm$0.006} & {0.327$\pm$0.024}      \\
3.7-4.3  &    1.64        &    4.12       & {-0.023$\pm$0.009} & {0.591$\pm$0.028}      \\
\hline
\end{tabular}
\end{table}

{
We note that 
there are positive slopes in Fig. 11 for the systems younger than
about 2.5-3 Gyr, in particular for the ranges $\log(l_-/l_0)=0-2.5$, $3.0-3.7$ and $3.7-4.3$.
The slope over all ranges, including $\log(l_-/l_0)=2.5-3.0$, is 0.0178.
There must be a heating mechanism responsible from the positive slope for ages less than 3 Gyr.
This mechanism is likely tidal interaction. It converts orbital (and spin, if any) energy of planets into heat.
This heat increases internal energy, and at the same time causes expansion. 
This part of Fig. 11 ($t< 3$ Gyr) is very consistent with Fig. 8. The maximum value of $\log(R_{\rm p}/R_{\rm j})$ in Fig. 8
is about 0.15. A careful consideration of Fig. 11 shows us that the slope for the younger systems gives
very similar increase in $\log(R_{\rm p}/R_{\rm j})$.
This implies that planetary orbit in some young systems is taken as circle, but
it may be eccentric. }

Storch \& Lai (2014) made model computations with different tidal dissipations.
They give results of two models (Models 1 and 2) in their fig. 5.
For Model 1 with low dissipation, radius inflation occur 
about 5 Gyr. Model 2, more dissipative than Model 1, however, has the highest radius 
about 2 Gyr. It seems that their Model 2 is more realistic than the other.

If we adopt that the variation in radius is a direct result of cooling and heating, then
one must answer why we see very different slopes for different $l_-$ intervals, 
for the systems with age greater than about 3 Gyr. 
{The answer might be that both cooling time and initial radius are very strong function of $l_-$.
According to Jupiter model constructed by Nettelmann et al. (2012),
its radius is initially 1.4 \RJ and drops 1.1 \RJ in 0.5 Gyr. 
That is to say the decrease in radius is 0.3 \RJ in the first  0.5 Gyr and 
0.1 \RJ in the following 4.2 Gyr (=4.7-0.5 Gyr). 
If $l_-$ is negligibly small then the cooling is very rapid and the radius reduces to \RJ in a relatively short 
time interval. 
If $l_-$ is not small then the cooling is so long that the time required for planet radius
to reduce 1 \RJ is much longer  than the MS lifetime of host star .}

If we exclude the systems younger we find the slope for the cooling part
as -0.0171.
\subsection{Variation of planetary radius with metallicity}
\begin{figure}
\includegraphics[width=85mm,angle=-90]{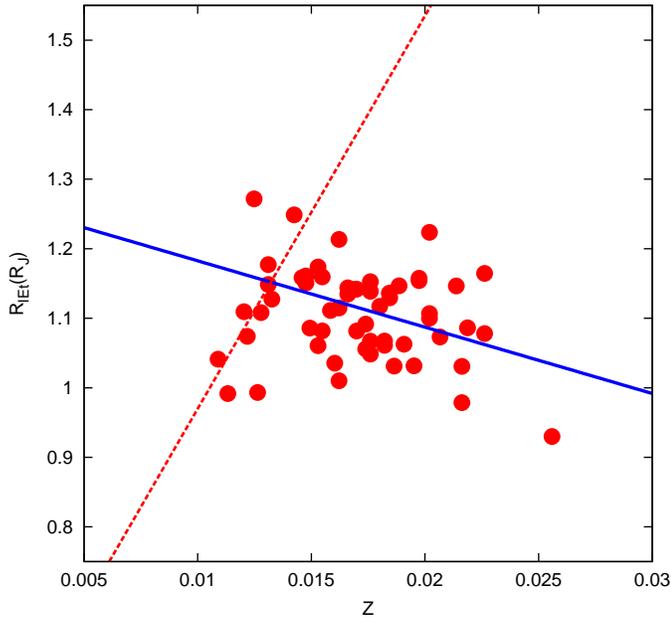}
\caption{ Planetary radius $R_{l\rm Et}$ (equation 25) with respect to $Z$. 
The planets with very large radii are excluded.
The only planets with $R_{\rm p} < 1.45 \RJ$ are considered. 
We subtract the effects of irradiation energy ($l_-$) and tidal interaction, 
and add the cooling effect to the observed $R_{\rm p}$. The resultant radius $R_{l\rm Et}$
is plotted \wrt $Z$. For the large part of the $Z$ range there is an anti-correlation between
$R_{l\rm Et}$ and $Z$. For low  $Z$, however, the relation between $R_{l\rm Et}$ and $Z$ is likely linear. 
The dashed and solid lines are for the ranges
$Z < 0.014$ and $Z > 0.014$, respectively.
}
\end{figure}

{
Radius of a planet depends on many parameters. 
These are irradiation energy ($l_-$), 
the tidal energy and the cooling rates. There are some studies in the literature
in which relation between planetary radius and stellar metallicity is examined (see, e.g. Guillot et al. (2006)  and 
Miller \& Fortney (2011)).
An inverse relation is found in these studies. Such a relation is very important for interior models of the planets.
We want to check if there is a relation for the majority of the up-to-date planetary data.

Effect of any parameter on the planetary radii can be subtracted from $R_{\rm p}$. For $l_-$, for example,
which is the most effective mechanism on $R_{\rm p}$,
\begin{equation}
R_l=R_{\rm p}-\frac{\Delta R_{\rm p}}{\Delta \log (l_-)} \delta \log (l_-).
\end{equation}
$R_l$ is the planetary radius if $l_-=0$.
After subtracting the effect of $l_-$, we find relation between $R_l$ and $\log(\dot{E})$.
Then, the effect of $\log(\dot{E})$ is also subtracted: 
\begin{equation}
R_{l \rm E}=R_{ l}-\frac{\Delta R_{ l}}{\Delta \log (\dot{E})} \delta \log (\dot{E}).
\end{equation}
For the effect of cooling,
\begin{equation}
R_{l\rm Et}=R_{ l\rm E}+\frac{\Delta R_{ l\rm E}}{\Delta t_9} \delta  t_9.
\end{equation}
$R_{l\rm Et}$ is the initial planetary radius  if there were no irradiation energy and tidal interaction.
In Fig. 12, $R_{l\rm Et}$ is plotted \wrt stellar metallicity $Z$.
We note that metallicity influences the planetary radius. 
For $Z>0.014$, the slope is negative and found as $-9.5\pm3.4$.
This result is in very good agreement with the findings of Guillot et al. (2006)  and
Miller \& Fortney (2011).
For $Z<0.014$, however, there is a very sharp positive slope for this narrow range. 
Thus, metallicity works on planetary radius in two opposite ways.  
The maximum radius occurs about $Z\approx 0.014$--$0.015$.
}

{
For our understanding of the relation  between metallicity and planet radius, two key parameters are very important.
They are mass of the metal core and opacity in the envelope. 
The metallicity is a very strong source of opacity. As metallicity increases, planet size does not rapidly decrease
because energy escape becomes difficult. 
This is the case for low-$Z$ range. The higher the metallicity is, the greater the radius is.
For the high metallicity regime, it is expected that mass of the metal core increases as metallicity increases.
The higher the metal core mass is, the smaller the radius is.
However, for this regime, the alternative explanation may be based on the high binding energies of heavy elements.
In the formation of planets, the majority of 
molecules is transformed into ionized gas, at least in the core regions. Some part of the heat gained from contraction 
is used for this transformation. If metallicity is very high then more energy is used to ionize matter and therefore
the interior of planets with high $Z$ becomes cooler than that of planets with low $Z$.
The cooler the interior is, the smaller the radius is.
}
 
\subsection{Variation of planetary mass in time and possible effect of evaporation}
{
To decipher detailed effects of irradiation on fundamental and orbital parameters of planets 
is a very difficult job. In addition to variation of planetary radius due to irradiation and time,
planetary mass may also change depending on the amount of energy the planets expose.   
\begin{figure}
\includegraphics[width=85mm,angle=-90]{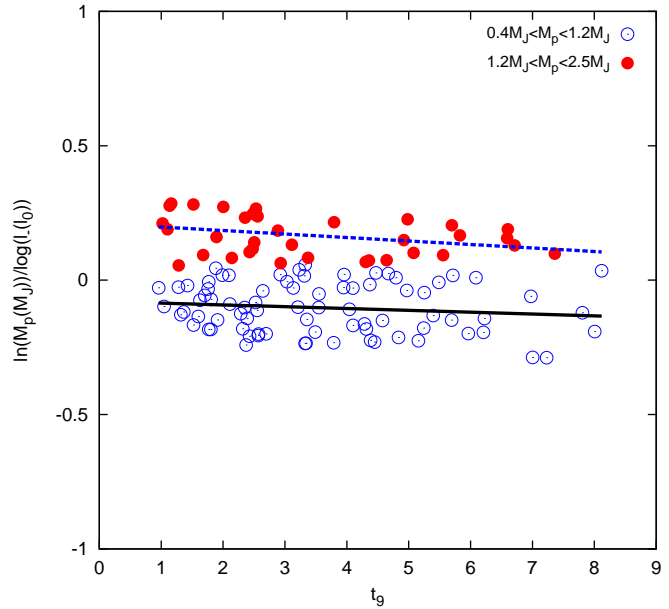}
\caption{  Planetary mass divided by $\log(l_-)$ with respect to age ($t_9$).
The solid and dashed lines are the fitted lines for the mass ranges 
$0.4\MJ < M_{\rm p}< 1.2 \MJ$ (circle) and $1.2\MJ < M_{\rm p}< 2.5 \MJ$ (filled circle), respectively.
}
\end{figure}
Planetary mass $\ln(M_{\rm p})$ divided by $\log(l_-)$ is plotted with respect to $t_9=t/(10^9$ Gyr$)$ in Fig. 13.
We note that there is a small, but not negligible, time variation of $\ln(M_{\rm p})/\log(l_-)$.
We also note that the slope for the young systems is higher than that for the old systems.
This is in agreement with the findings of Valencia et al. (2010) and Lopez, Fortney \& Miller (2012).
From the slope over all range of age, we can write
\begin{equation}
\frac{\Delta (\ln(M_{\rm p})/\log(l_-))}{\Delta t_9}=-{(0.0130\pm0.0064)}.
\end{equation}
If we assume that time variation of $l_-$ is negligibly small, we find
\begin{equation}
\dot{M_{\rm p}}=\frac{\Delta \ln(M_{\rm p})}{\Delta t_9}=-{(0.0130\pm0.0064)} \log(l_-).
\end{equation}
for $1.2\MJ < M_{\rm p}< 2.5 \MJ$.
This variation of planetary mass may be a result of evaporation.
Equation (27) gives mass-loss rate for these giant gas planets as $10^{11}-10^{13}$ g s$^{-1}$.
The slope in Fig. 13 for the lower planetary mass range ($0.4\MJ < M_{\rm p}< 1.2 \MJ$), 
however, is $0.0068\pm0.0058$. It is significantly less than 0.013. 
Then, for the low-mass giant gas planets ($M_{\rm p}< 1.2 \MJ$)
\begin{equation}
\dot{M_{\rm p}}=\frac{\Delta \ln(M_{\rm p})}{\Delta t_9}=-{(0.0068\pm0.0058)} \log(l_-).
\end{equation}
This gives mass-loss rates of the giants gas planets as 
$10^{10}$--$10^{12}$ g s$^{-1}$.

There are several studies in the literature devoted to present 
behaviour of the planetary mass-loss rates (see, e.g Valencia et al. 2010 and Lopez, Fortney \& Miller 2012).
Linsky et al. (2010) find mass-loss rate for HD209458 b as $(0.8$--$4) 10^{11}$ g s$^{-1}$. 
Using its age and $\log(l_-)$, given in the online Table A1, the mass-loss rate is found as  
$\dot{M_{\rm p}}=  (1$--$18) 10^{11}$ g s$^{-1}$ from equation (28).
Although its uncertainty is high, 
the lower part of its range is in agreement with the observed rate.

}
\subsection{Initial planetary mass}
If evaporation of planets really occurs, we can estimate the initial masses ($M_{\rm pi}$) of the giant gas planets
{by using equation (27) and (28):
for $1.2\MJ < M_{\rm p}< 2.5 \MJ$,
\begin{equation}
M_{\rm pi}=M_{\rm p}{\rm e}^{{(0.0130\pm0.0064)}\log(l_-)t_9};
\end{equation}
for $0.4\MJ < M_{\rm p}< 1.2 \MJ$,
\begin{equation}
M_{\rm pi}=M_{\rm p}{\rm e}^{{(0.0068\pm0.0058)}\log(l_-)t_9}.
\end{equation}

According to our findings, the mean value of the mass lost by the giant gas planets is 
about 12 per cent and  the most irradiated planets lose 5 per cent of their mass in every 1 Gyr. 
The maximum amount of lost mass is about 33 per cent (HATS-2 b).
These preliminary results need further investigation and must be tested.
}


\section{Conclusion}
There are many papers in the literature on the effect of the induced flux on the radius of planets. 
The main outcome of these studies is that the radius is greater in the systems with the high incident flux than
in the systems with the low flux.
The excess depends on the mass interval of the planet. We show that there is much more definite relation 
between radius and energy per gram per second ($l_-$), if $l_-$ is greater than $300-500$ $l_{0}$,
where is irradiation energy per unit mass and time by a planet with 1 \MJ and 1 \RJ at 1 au in our Solar system. 
This mechanism is the most efficient one on planetary size.
If $l_-$ is greater than 
3000 $l_{0}$, then there is another relation. $T_{\rm eq}$ of these planets in this range is higher
than 1500 K. The reason for this extra inflation may be due to dissociation of molecules. It is demonstrated in Fig. 10
that there are some giant planets, which  have material in the atomic form in the heated region rather than molecular, and 
excess in radius.
The maximum effect of molecular dissociation on radius is about {33} per cent.

Tidal interaction influences radius of giant gas planets if the orbit is eccentric. 
Using expression for energy rate given by  Storch \& Lai (2014), we show that this mechanism
causes giant gas planets to expand up to  15 per cent and confirm that their Model 2 seems more realistic than Model 1.

We also develop new methods to determine metallicity and age of the host stars. If metallicity is computed 
from direct use of observed [Fe/H] values, $Z_{\rm Fe}$ ranges from 0.006 to 0.05. 
Regarding the facts that there is an inverse relation between [O/Fe] and [Fe/H] (Edvardsson et al. 1993) and 
oxygen is the most abundant heavy element, we also compute $Z_{\rm O}$ from [O/H] by using the observed relation 
between [O/H] and [Fe/H]. $Z_{\rm O}$ ranges from 0.009 to 0.026. The difference between $Z_{\rm O}$ and $Z_{\rm Fe}$
is very significant for high metallicity.

A new method is developed to compute age of the host stars. The method is based on 
their radius, mass and metallicity. It is relatively easy to apply.
By using $Z_{\rm O}$, we find that age of the host stars ranges from 0.3 to 11.1 Gyr.
The mean age is 4.2 Gyr. If we use $Z_{\rm Fe}$ as metallicity of the host stars,
age is found as greater than the galactic age. This implies that $Z_{\rm O}$ is a better 
indicator for the total metallicity. Our method yields results in very good agreement with 
the ages from asteroseismic inferences (see Table 1).

We also present our preliminary results for time variation of planetary radius due to cooling 
and mass due to evaporation.
The mass-loss 
is also a function of irradiation energy per gram per second. We find that highly irradiated 
gas planets ($\log(l_-/l_0)> 3.2$) loss about 5 per cent of their total mass in every 1 Gyr.

{
We also test if metallicity has any influence on planetary radius. 
We subtract the effects of irradiation energy, tidal energy rate and cooling on the radius.
The resultant radius shows an inverse relation with stellar metallicity. For the low-metallicity regime,
there are less data, but the situation is opposite of this. }

\section*{Acknowledgements}
{The anonymous referee is acknowledged
for suggestions which improved the presentation of the manuscript.}
This work is supported by the Scientific and Technological Research
Council of Turkey (T\"UB\.ITAK: 112T989).

\newpage
\onecolumn
\appendix
\section[]{Online-only table for basic properties of the giant gas planets and their host stars  }
\small\addtolength{\tabcolsep}{-4pt}
\begin{longtable}{@{}lrcccclccccccrr@{}}
\caption[Continued.]{
Fundamental properties of the giant gas planets and their host stars.
The data are taken from TEPCat. Columns are organized as name, mass, radius and effective temperature of the host stars,
semi-major axis, period, eccentricity, planetary mass, radius, equilibrium temperature, incident flux, energy 
received per unit mass per unit time, and rate of energy dissipation due to tidal interaction. In the last two columns, 
age and metallicity computed from stellar properties are listed (see Section 3).
}\\
\hline
Planet               & $M$ &      $R$    &   $T_{\rm eff}$ &       $a$ &       $P$ &       $e$  &       $M_{\rm p}$ &       $R_{\rm p}$  &       $T_{\rm eq}$ &       $F_{\rm I}$   &       $l_-$   &       $\log(\dot{E})$   &      age  & $Z_{\rm O} $       \\
        & (M$_{\sun}$) &     (R$_{\sun}$) &       (K)   &       (AU)   &       (d)     &      &       (M$_{\rm j}$)   &       (R$_{\rm j})$   &       (K)  & (${\rm F}_{\rm \oplus})$&       ($l_{0})$   &   (erg s$^{-1}$)    &(Gyr)&    \\
\hline
\endfirsthead

\caption[-- continued from previous page]{-- continued from previous page}\\
\hline
Planet    & $M$ &      $R$    &   $T_{\rm eff}$ &       $a$ &       $P$ &       $e$  &       $M_{\rm p}$ &       $R_{\rm p}$  &       $T_{\rm eq}$ &       ${\rm F}_{\rm I}$   &       $l_-$   &       $\log(\dot{E})$   &      age  & $Z_{\rm O} $       \\
        & (M$_{\sun}$) &     (R$_{\sun}$) &       (K)   &       (AU)   &       (d)     &      &       (M$_{\rm j}$)   &       (R$_{\rm j})$   &       (K)  & (${\rm F}_{\rm \oplus})$&       ($l_{0})$   &   (erg s$^{-1}$)    &(Gyr)&    \\

\hline
\endhead

\hline
\endfoot

CoRoT-01  b& 0.95& 1.13& 5950& 0.0254&  1.509& 0.000& 1.03& 1.55& 1915&   2238$\pm$    647&   5227$\pm$    938&    --- &  6.7$\pm$  3.2&  0.011$\pm$  0.003\\
CoRoT-02  b& 1.00& 0.90& 5598& 0.0283&  1.743& 0.014& 3.57& 1.46& 1521&    890$\pm$     98&    531$\pm$     44&    26.5$\pm$     0.3&  1.1$\pm$  1.6&  0.016$\pm$  0.004\\
CoRoT-03  b& 1.40& 1.58& 6740& 0.0578&  4.257& 0.000&21.96& 1.04& 1695&   1374$\pm$    330&     67$\pm$     11&    --- &  1.1$\pm$  0.3&  0.015$\pm$  0.004\\
CoRoT-04  b& 1.19& 1.15& 6190& 0.0912&  9.202& 0.000& 0.73& 1.16& 1058&    208$\pm$     47&    384$\pm$    113&    --- &  1.1$\pm$  0.6&  0.016$\pm$  0.004\\
CoRoT-05  b& 1.02& 1.05& 6100& 0.0500&  4.038& 0.090& 0.47& 1.18& 1348&    549$\pm$    143&   1635$\pm$    480&    25.5$\pm$     0.8&  3.3$\pm$  2.1&  0.011$\pm$  0.003\\
CoRoT-06  b& 1.05& 1.04& 6090& 0.0855&  8.887& 0.000& 2.96& 1.18& 1025&    183$\pm$     26&     87$\pm$     16&    --- &  2.5$\pm$  1.3&  0.012$\pm$  0.003\\
CoRoT-09  b& 0.96& 0.94& 5613& 0.4027& 95.274& 0.110& 0.83& 1.04&  413&      4$\pm$      1&      6$\pm$      1&    17.1$\pm$     0.7&  3.8$\pm$  2.8&  0.015$\pm$  0.004\\
CoRoT-10  b& 0.90& 0.74& 5075& 0.1060& 13.241& 0.530& 2.78& 0.94&  647&     29$\pm$      6&      9$\pm$      2&    23.0$\pm$     0.6&     ---   &  0.021$\pm$  0.005\\
CoRoT-11  b& 1.26& 1.37& 6440& 0.0440&  2.994& 0.000& 2.34& 1.43& 1735&   1505$\pm$    355&   1308$\pm$    322&    --- &  2.0$\pm$  1.1&  0.015$\pm$  0.004\\
CoRoT-12  b& 1.02& 1.05& 5675& 0.0394&  2.828& 0.000& 0.89& 1.35& 1410&    656$\pm$    127&   1348$\pm$    268&    --- &  5.0$\pm$  2.7&  0.018$\pm$  0.005\\
CoRoT-13  b& 1.09& 1.27& 5945& 0.0510&  4.035& 0.000& 1.31& 1.25& 1432&    699$\pm$    163&    836$\pm$    162&    --- &  5.6$\pm$  1.6&  0.015$\pm$  0.004\\
CoRoT-14  b& 1.13& 1.19& 6035& 0.0269&  1.512& 0.000& 7.67& 1.02& 1936&   2335$\pm$    840&    315$\pm$     69&    --- &  3.4$\pm$  1.6&  0.016$\pm$  0.004\\
CoRoT-15  b& 1.31& 1.36& 6350& 0.0458&  3.060& 0.000&64.90& 1.04& 1670&   1287$\pm$   1001&     21$\pm$     15&    --- &  1.3$\pm$  1.0&  0.017$\pm$  0.004\\
CoRoT-16  b& 1.10& 1.19& 5650& 0.0618&  5.352& 0.330& 0.54& 1.17&  --&    339$\pm$    120&    867$\pm$    374&    26.1$\pm$     1.0&  4.8$\pm$  1.7&  0.019$\pm$  0.005\\
CoRoT-17  b& 1.04& 1.62& 5740& 0.0481&  3.768& 0.000& 2.46& 1.01& 1610&   1105$\pm$    754&    455$\pm$    359&    --- &  7.4$\pm$  1.8&  0.015$\pm$  0.004\\
CoRoT-18  b& 0.86& 0.92& 5440& 0.0286&  1.900& 0.000& 3.27& 1.25& 1490&    820$\pm$    200&    392$\pm$     72&    --- &  7.5$\pm$  3.7&  0.014$\pm$  0.003\\
CoRoT-19  b& 1.18& 1.58& 6090& 0.0512&  3.897& 0.000& 1.09& 1.19& 1630&   1170$\pm$    661&   1520$\pm$    793&    --- &  4.5$\pm$  1.4&  0.015$\pm$  0.004\\
CoRoT-20  b& 1.11& 1.34& 5880& 0.0892&  9.243& 0.000& 5.06& 1.16& 1100&    242$\pm$    163&     64$\pm$     33&    --- &  6.2$\pm$  1.8&  0.018$\pm$  0.005\\
CoRoT-21  b& 1.39& 1.95& 6200& 0.0417&  2.725& 0.000& 2.26& 1.27&  --&   2901$\pm$    965&   2070$\pm$    758&    --- &  2.5$\pm$  0.4&  0.015$\pm$  0.004\\
CoRoT-23  b& 1.12& 1.74& 5900& 0.0481&  3.631& 0.160& 3.06& 1.18& 1710&   1424$\pm$    598&    648$\pm$    276&    26.3$\pm$     1.1&  5.7$\pm$  1.5&  0.016$\pm$  0.004\\
CoRoT-26  b& 1.09& 1.79& 5590& 0.0526&  4.205& 0.000& 0.52& 1.26& 1600&   1015$\pm$    315&   3099$\pm$    937&    --- &  6.3$\pm$  1.2&  0.015$\pm$  0.004\\
HAT-P-01  b& 1.15& 1.17& 5975& 0.0556&  4.465& 0.000& 0.52& 1.32& 1322&    510$\pm$     54&   1690$\pm$    109&    --- &  2.7$\pm$  0.3&  0.018$\pm$  0.004\\
HAT-P-02  b& 1.28& 1.68& 6290& 0.0674&  5.633& 0.508$^a$& 8.74& 1.19& 1516&    873$\pm$    210&    141$\pm$     32&    26.3$\pm$     0.6&  3.7$\pm$  0.4&  0.018$\pm$  0.005\\
HAT-P-03  b& 0.90& 0.87& 5185& 0.0384&  2.900& 0.000& 0.58& 0.95& 1189&    332$\pm$     42&    510$\pm$     55&    --- &  6.0$\pm$  3.6&  0.021$\pm$  0.005\\
HAT-P-04  b& 1.27& 1.60& 5860& 0.0446&  3.057& 0.004$^a$& 0.68& 1.34& 1691&   1359$\pm$    356&   3573$\pm$    663&    23.7$\pm$     0.7&  4.0$\pm$  1.1&  0.020$\pm$  0.005\\
HAT-P-05  b& 1.16& 1.14& 5960& 0.0408&  2.788& 0.000& 1.06& 1.25& 1517&    880$\pm$    144&   1301$\pm$    224&    --- &  2.0$\pm$  1.1&  0.020$\pm$  0.005\\
HAT-P-06  b& 1.29& 1.52& 6570& 0.0524&  3.853& 0.023$^a$& 1.06& 1.39& 1704&   1401$\pm$    248&   2566$\pm$    435&    24.7$\pm$     0.5&  2.1$\pm$  0.5&  0.013$\pm$  0.003\\
HAT-P-07  b& 1.51& 1.96& 6350& 0.0380&  2.205& 0.005$^a$& 1.80& 1.47& 2194&   3853$\pm$    345&   4597$\pm$    202&    25.1$\pm$     0.2&  1.9$\pm$  0.1&  0.021$\pm$  0.005\\
HAT-P-08  b& 1.19& 1.48& 6200& 0.0439&  3.076& 0.003$^a$& 1.27& 1.32& 1713&   1497$\pm$    208&   2049$\pm$    209&    23.4$\pm$     0.4&  4.4$\pm$  0.9&  0.015$\pm$  0.004\\
HAT-P-09  b& 1.28& 1.34& 6350& 0.0529&  3.923& 0.000& 0.78& 1.38& 1540&    935$\pm$    249&   2289$\pm$    575&    --- &  1.6$\pm$  0.8&  0.018$\pm$  0.004\\
HAT-P-13  b& 1.32& 1.76& 5653& 0.0438&  2.916& 0.013$^a$& 0.91& 1.49& 1725&   1471$\pm$    216&   3591$\pm$    316&    25.1$\pm$     0.3&  3.9$\pm$  0.6&  0.024$\pm$  0.006\\
HAT-P-14  b& 1.42& 1.59& 6600& 0.0611&  4.628& 0.115$^a$& 2.27& 1.22& 1624&   1155$\pm$    173&    756$\pm$    100&    25.4$\pm$     0.4&  1.2$\pm$  0.3&  0.017$\pm$  0.004\\
HAT-P-15  b& 1.01& 1.08& 5568& 0.0964& 10.864& 0.208$^a$& 1.95& 1.07&  904&    108$\pm$     17&     63$\pm$      7&    23.5$\pm$     0.4&  6.4$\pm$  1.5&  0.020$\pm$  0.005\\
HAT-P-16  b& 1.22& 1.16& 6140& 0.0413&  2.776& 0.042$^a$& 4.19& 1.19& 1567&   1003$\pm$    120&    338$\pm$     31&    25.8$\pm$     0.3&  0.8$\pm$  0.7&  0.018$\pm$  0.004\\
HAT-P-17  b& 0.86& 0.84& 5246& 0.0882& 10.339& 0.342$^a$& 0.53& 1.01&  792&     61$\pm$      8&    117$\pm$     10&    24.1$\pm$     0.3&  5.3$\pm$  2.8&  0.015$\pm$  0.004\\
HAT-P-20  b& 0.76& 0.69& 4595& 0.0361&  2.876& 0.016$^a$& 7.25& 0.87&  970&    147$\pm$     23&     15$\pm$      1&    24.1$\pm$     0.3&  5.6$\pm$  4.9&  0.023$\pm$  0.006\\
HAT-P-21  b& 0.95& 1.11& 5588& 0.0494&  4.124& 0.228& 4.06& 1.02& 1283&    438$\pm$    103&    113$\pm$     24&    26.0$\pm$     0.6&  8.4$\pm$  1.8&  0.015$\pm$  0.004\\
HAT-P-22  b& 0.92& 1.04& 5302& 0.0414&  3.212& 0.006$^a$& 2.15& 1.08& 1283&    447$\pm$     75&    243$\pm$     33&    23.5$\pm$     0.4& 11.1$\pm$  2.1&  0.020$\pm$  0.005\\
HAT-P-23  b& 1.13& 1.20& 5905& 0.0232&  1.213& 0.106& 2.09& 1.37& 2056&   2935$\pm$    570&   2628$\pm$    485&    29.1$\pm$     0.4&  3.8$\pm$  0.7&  0.018$\pm$  0.005\\
HAT-P-24  b& 1.19& 1.29& 6373& 0.0464&  3.355& 0.033$^a$& 0.68& 1.24& 1624&   1151$\pm$    210&   2612$\pm$    421&    25.2$\pm$     0.4&  2.6$\pm$  0.5&  0.013$\pm$  0.003\\
HAT-P-25  b& 1.01& 0.96& 5500& 0.0466&  3.653& 0.032& 0.57& 1.19& 1202&    347$\pm$     66&    868$\pm$    151&    24.8$\pm$     0.5&  3.5$\pm$  1.2&  0.022$\pm$  0.005\\
HAT-P-27  b& 0.94& 0.90& 5316& 0.0403&  3.040& 0.078& 0.66& 1.04& 1207&    356$\pm$     66&    581$\pm$    115&    25.8$\pm$     0.5&  4.6$\pm$  1.9&  0.022$\pm$  0.005\\
HAT-P-28  b& 1.02& 1.10& 5680& 0.0434&  3.257& 0.051& 0.63& 1.21& 1384&    603$\pm$    157&   1416$\pm$    347&    25.6$\pm$     0.7&  5.7$\pm$  1.4&  0.018$\pm$  0.004\\
HAT-P-29  b& 1.21& 1.22& 6087& 0.0667&  5.723& 0.061$^a$& 0.78& 1.11& 1260&    415$\pm$    124&    653$\pm$    224&    24.1$\pm$     0.8&  2.1$\pm$  0.6&  0.020$\pm$  0.005\\
HAT-P-30  b& 1.24& 1.22& 6338& 0.0419&  2.811& 0.020$^a$& 0.71& 1.34& 1630&   1218$\pm$    163&   3076$\pm$    419&    25.4$\pm$     0.4&  1.0$\pm$  0.5&  0.018$\pm$  0.004\\
HAT-P-31  b& 1.22& 1.36& 6065& 0.0550&  5.005& 0.242$^a$& 2.17& 1.07& 1450&    742$\pm$    749&    391$\pm$    194&    25.9$\pm$     2.9&  3.2$\pm$  1.0&  0.018$\pm$  0.005\\
HAT-P-32  b& 1.16& 1.22& 6207& 0.0343&  2.150& 0.200$^a$& 0.86& 1.79& 1786&   1683$\pm$    178&   6264$\pm$   1369&    28.8$\pm$     0.2&  2.6$\pm$  0.6&  0.015$\pm$  0.004\\
HAT-P-33  b& 1.38& 1.64& 6446& 0.0499&  3.475& 0.130$^a$& 0.76& 1.69& 1782&   1668$\pm$    193&   6223$\pm$   1157&    27.0$\pm$     0.3&  1.8$\pm$  0.2&  0.017$\pm$  0.004\\
HAT-P-34  b& 1.39& 1.53& 6442& 0.0677&  5.453& 0.411$^a$& 3.33& 1.20& 1520&    794$\pm$    202&    342$\pm$     94&    26.4$\pm$     0.7&  1.4$\pm$  0.3&  0.020$\pm$  0.005\\
HAT-P-35  b& 1.24& 1.43& 6178& 0.0498&  3.647& 0.025& 1.05& 1.33& 1581&   1085$\pm$    184&   1828$\pm$    326&    24.9$\pm$     0.5&  3.3$\pm$  0.5&  0.018$\pm$  0.004\\
HAT-P-36  b& 1.02& 1.10& 5560& 0.0238&  1.327& 0.063& 1.83& 1.26& 1823&   1819$\pm$    377&   1586$\pm$    263&    28.2$\pm$     0.5&  6.6$\pm$  1.7&  0.021$\pm$  0.005\\
HAT-P-37  b& 0.93& 0.88& 5500& 0.0379&  2.797& 0.058& 1.17& 1.18& 1271&    439$\pm$    105&    522$\pm$    114&    26.0$\pm$     0.5&  3.3$\pm$  2.1&  0.016$\pm$  0.004\\
HAT-P-39  b& 1.40& 1.63& 6340& 0.0509&  3.544& 0.000& 0.60& 1.57& 1752&   1478$\pm$    275&   6091$\pm$   1844&    --- &  1.6$\pm$  0.3&  0.019$\pm$  0.005\\
HAT-P-40  b& 1.51& 2.21& 6080& 0.0608&  4.457& 0.000& 0.62& 1.73& 1770&   1615$\pm$    227&   7860$\pm$   1049&    --- &  2.3$\pm$  0.2&  0.020$\pm$  0.005\\
HAT-P-41  b& 1.42& 1.68& 6390& 0.0426&  2.694& 0.000& 0.80& 1.68& 1941&   2336$\pm$    362&   8291$\pm$   1805&    --- &  1.7$\pm$  0.3&  0.020$\pm$  0.005\\
HAT-P-42  b& 1.18& 1.53& 5743& 0.0575&  4.642& 0.000& 0.98& 1.28& 1427&    689$\pm$    173&   1153$\pm$    418&    --- &  5.5$\pm$  1.0&  0.021$\pm$  0.005\\
HAT-P-43  b& 1.05& 1.10& 5645& 0.0443&  3.333& 0.000& 0.66& 1.28& 1361&    566$\pm$     76&   1412$\pm$    303&    --- &  5.4$\pm$  0.9&  0.020$\pm$  0.005\\
HAT-P-45  b& 1.26& 1.32& 6330& 0.0452&  3.129& 0.049& 0.89& 1.43& 1652&   1227$\pm$    404&   2798$\pm$   1116&    26.0$\pm$     0.8&  1.8$\pm$  0.6&  0.017$\pm$  0.004\\
HAT-P-46  b& 1.28& 1.40& 6120& 0.0577&  4.463& 0.123& 0.49& 1.28& 1458&    737$\pm$    393&   2465$\pm$   1450&    25.7$\pm$     1.3&  2.4$\pm$  0.9&  0.022$\pm$  0.005\\
HATS-1    b& 0.99& 1.04& 5870& 0.0444&  3.446& 0.120& 1.86& 1.30& 1359&    582$\pm$    204&    532$\pm$    207&    26.4$\pm$     0.8&  5.0$\pm$  1.7&  0.014$\pm$  0.004\\
HATS-2    b& 0.88& 0.90& 5227& 0.0230&  1.354& 0.000& 1.35& 1.17& 1577&   1021$\pm$    144&   1036$\pm$    168&    --- &  7.4$\pm$  2.5&  0.018$\pm$  0.005\\
HATS-3    b& 1.21& 1.40& 6351& 0.0485&  3.548& 0.000& 1.07& 1.38& 1648&   1224$\pm$    131&   2179$\pm$    387&    --- &  2.9$\pm$  0.4&  0.013$\pm$  0.003\\
HD017156 b& 1.30& 1.49& 6079& 0.1637& 21.216& 0.675& 3.26& 1.07&  883&    101$\pm$     13&     35$\pm$      3&         ---     &  2.7$\pm$  0.6&  0.020$\pm$  0.005\\
HD080606 b& 1.02& 1.04& 5574& 0.4564&111.437& 0.933& 4.11& 1.00&  405&      4$\pm$      0&      1$\pm$      0&         ---     &  5.8$\pm$  2.0&  0.023$\pm$  0.006\\
HD189733 b& 0.84& 0.75& 5050& 0.0314&  2.219& 0.004& 1.15& 1.15& 1191&    334$\pm$     46&    385$\pm$     38&    24.3$\pm$     0.4&  1.6$\pm$  3.2&  0.015$\pm$  0.004\\
HD209458 b& 1.15& 1.16& 6117& 0.0475&  3.525& 0.000& 0.71& 1.38& 1459&    753$\pm$     60&   2008$\pm$     97&    --- &  2.3$\pm$  0.6&  0.016$\pm$  0.004\\
KELT-1    b& 1.34& 1.47& 6516& 0.0247&  1.218& 0.010&27.38& 1.12& 2423&   5731$\pm$    703&    260$\pm$     26&    26.5$\pm$     0.4&  1.5$\pm$  0.4&  0.016$\pm$  0.004\\
KELT-2    b& 1.31& 1.84& 6151& 0.0550&  4.114& 0.000& 1.52& 1.29& 1712&   1430$\pm$    192&   1561$\pm$    245&    --- &  3.1$\pm$  0.5&  0.016$\pm$  0.004\\
KELT-3    b& 1.28& 1.47& 6306& 0.0412&  2.703& 0.000& 1.48& 1.35& 1811&   1810$\pm$    275&   2217$\pm$    336&    --- &  2.5$\pm$  0.5&  0.016$\pm$  0.004\\
KELT-6    b& 1.09& 1.58& 6102& 0.0794&  7.846& 0.220& 0.43& 1.19& 1313&    493$\pm$    126&   1631$\pm$    526&    24.6$\pm$     0.7&  5.2$\pm$  0.7&  0.011$\pm$  0.003\\
Kepler-05 b& 1.30& 1.54& 6297& 0.0497&  3.548& 0.000& 2.03& 1.21& 1692&   1364$\pm$    177&    980$\pm$     80&    --- &  2.6$\pm$  0.3&  0.016$\pm$  0.004\\
Kepler-06 b& 1.11& 1.26& 5647& 0.0444&  3.235& 0.000& 0.63& 1.17& 1451&    737$\pm$    162&   1591$\pm$    305&    --- &  6.2$\pm$  2.9&  0.023$\pm$  0.006\\
Kepler-07 b& 1.41& 2.03& 5933& 0.0632&  4.885& 0.000& 0.45& 1.65& 1619&   1144$\pm$    117&   6867$\pm$   1347&    --- &  2.6$\pm$  0.5&  0.017$\pm$  0.004\\
Kepler-08 b& 1.23& 1.50& 6213& 0.0485&  3.523& 0.000& 0.59& 1.38& 1662&   1271$\pm$    248&   4108$\pm$   1055&    --- &  3.4$\pm$  0.7&  0.014$\pm$  0.004\\
Kepler-12 b& 1.16& 1.49& 5947& 0.0555&  4.438& 0.000& 0.43& 1.71& 1485&    809$\pm$    184&   5478$\pm$   1118&    --- &  5.2$\pm$  1.9&  0.017$\pm$  0.004\\
Kepler-14 b& 1.32& 2.09& 6395& 0.0771&  6.790& 0.040& 7.68& 1.13& 1605&   1103$\pm$    189&    182$\pm$     24&    23.3$\pm$     0.4&  3.3$\pm$  0.5&  0.018$\pm$  0.004\\
Kepler-15 b& 1.08& 1.25& 5595& 0.0583&  4.943& 0.000& 0.70& 1.29& 1251&    406$\pm$    100&    970$\pm$    219&    --- &  7.8$\pm$  3.0&  0.023$\pm$  0.006\\
Kepler-17 b& 1.07& 0.98& 5781& 0.0260&  1.486& 0.000& 2.34& 1.31& 1712&   1427$\pm$    184&   1047$\pm$     78&    --- &  1.5$\pm$  1.4&  0.021$\pm$  0.005\\
Kepler-30 c& 0.99& 0.95& 5498& 0.3000& 60.323& 0.011& 2.01& 1.10&   --&      8$\pm$    -52&      4$\pm$      0&    16.4$\pm$   -19.7&  3.6$\pm$  2.8&  0.019$\pm$  0.005\\
Kepler-39 b& 1.08& 1.23& 6260& 0.1539& 21.087& 0.121&17.90& 1.09&  851&     87$\pm$     29&      5$\pm$      1&    21.2$\pm$     0.9&  4.0$\pm$  2.2&  0.011$\pm$  0.003\\
Kepler-40 b& 1.46& 2.48& 6510& 0.0802&  6.873& 0.000& 2.16& 1.44& 1744&   1541$\pm$    504&   1480$\pm$    541&    --- &  2.3$\pm$  0.6&  0.017$\pm$  0.004\\
Kepler-41 b& 0.94& 0.94& 5660& 0.0290&  1.856& 0.000& 0.49& 0.85& 1554&    969$\pm$    258&   1427$\pm$    446&    --- &  4.4$\pm$  3.4&  0.014$\pm$  0.003\\
Kepler-43 b& 1.24& 1.33& 6041& 0.0440&  3.024& 0.000& 3.09& 1.12& 1603&   1095$\pm$    244&    440$\pm$     62&    --- &  2.9$\pm$  1.2&  0.022$\pm$  0.006\\
Kepler-44 b& 1.21& 1.46& 5757& 0.0457&  3.247& 0.000& 1.03& 1.20& 1568&   1006$\pm$    334&   1407$\pm$    376&    --- &  4.8$\pm$  1.5&  0.021$\pm$  0.005\\
Kepler-56 c& 1.32& 4.23& 4840& 0.1652& 21.402&-1.000& 0.57& 0.87&  --&    323$\pm$     71&    433$\pm$     90&         ---     &  3.4$\pm$  0.9&  0.019$\pm$  0.005\\
Kepler-74 b& 1.40& 1.51& 6050& 0.0840&  7.341& 0.287& 0.68& 1.32& 1250&    388$\pm$    229&    995$\pm$    342&    25.3$\pm$     1.7&  1.3$\pm$  0.7&  0.023$\pm$  0.006\\
Kepler-75 b& 0.88& 0.88& 5330& 0.0800&  8.885& 0.569& 9.90& 1.03&  850&     87$\pm$     26&      9$\pm$      1&         ---     &  5.3$\pm$  3.4&  0.014$\pm$  0.004\\
Kepler-77 b& 0.95& 0.99& 5520& 0.0450&  3.579& 0.000& 0.43& 0.96& 1440&    403$\pm$     45&    864$\pm$     93&    --- &  7.0$\pm$  1.9&  0.019$\pm$  0.005\\
Kepler-87 b& 1.10& 1.82& 5600& 0.4810&114.736& 0.036& 1.02& 1.20&  478&     12$\pm$      2&     17$\pm$      1&    15.9$\pm$     0.6&  5.3$\pm$  0.9&  0.012$\pm$  0.003\\
Kepler-91 b& 1.31& 6.20& 4550& 0.0720&  6.247& 0.066& 0.88& 1.38&  --&   2853$\pm$    493&   6210$\pm$   1298&    24.4$\pm$     0.4&  3.3$\pm$  0.7&  0.017$\pm$  0.004\\
KOI-205   b& 0.93& 0.84& 5237& 0.0987& 11.720& 0.000&39.90& 0.81&  737&     49$\pm$      5&      0$\pm$      0&    --- &  2.2$\pm$  1.8&  0.018$\pm$  0.005\\
KOI-415   b& 0.94& 1.25& 5810& 0.5930&166.788& 0.698&62.14& 0.79&  --&      4$\pm$      0&      0$\pm$      0&         ---     &  9.3$\pm$  2.1&  0.011$\pm$  0.003\\
OGLE-TR-010 b& 1.28& 1.52& 6075& 0.0452&  3.101& 0.000& 0.68& 1.72& 1702&   1385$\pm$    322&   6027$\pm$   2100&    --- &  3.5$\pm$  0.8&  0.021$\pm$  0.005\\
OGLE-TR-056 b& 1.34& 1.74& 6119& 0.0245&  1.212& 0.000& 1.41& 1.73& 2482&   6311$\pm$    896&  13458$\pm$   2522&    --- &  3.4$\pm$  0.7&  0.020$\pm$  0.005\\
OGLE-TR-111 b& 0.85& 0.82& 5044& 0.0468&  4.014& 0.000& 0.55& 1.01& 1019&    179$\pm$     31&    333$\pm$     85&    --- &  6.6$\pm$  5.3&  0.019$\pm$  0.005\\
OGLE-TR-113 b& 0.75& 0.77& 4790& 0.0226&  1.432& 0.000& 1.23& 1.09& 1342&    540$\pm$    135&    520$\pm$    136&    --- &  9.7$\pm$ 11.7&  0.017$\pm$  0.004\\
OGLE-TR-132 b& 1.29& 1.34& 6210& 0.0303&  1.690& 0.000& 1.17& 1.23& 1991&   2610$\pm$    571&   3370$\pm$    843&    --- &  1.9$\pm$  1.1&  0.023$\pm$  0.006\\
OGLE-TR-182 b& 1.19& 1.53& 5924& 0.0520&  3.979& 0.000& 1.06& 1.47& 1550&    955$\pm$    277&   1947$\pm$    646&    --- &  5.7$\pm$  0.7&  0.023$\pm$  0.006\\
OGLE-TR-211 b& 1.31& 1.56& 6325& 0.0510&  3.677& 0.000& 0.75& 1.26& 1686&   1341$\pm$    426&   2849$\pm$   1283&    --- &  2.5$\pm$  0.5&  0.017$\pm$  0.004\\
OGLE-TR-L9 b& 1.43& 1.50& 6933& 0.0405&  2.486& 0.000& 4.40& 1.63& 2034&   2845$\pm$    393&   1724$\pm$    684&    --- &  0.5$\pm$  0.4&  0.014$\pm$  0.004\\
Qatar-1   1b& 0.85& 0.80& 4910& 0.0234&  1.420& 0.000& 1.33& 1.18& 1389&    608$\pm$    131&    636$\pm$    120&    --- &  5.1$\pm$  2.6&  0.019$\pm$  0.005\\
TrES-1    b& 0.89& 0.82& 5226& 0.0395&  3.030& 0.000& 0.76& 1.10& 1147&    287$\pm$     36&    456$\pm$     59&    --- &  2.7$\pm$  2.9&  0.016$\pm$  0.004\\
TrES-2    b& 0.99& 0.96& 5850& 0.0357&  2.471& 0.004$^a$& 1.21& 1.19& 1466&    768$\pm$     83&    906$\pm$     71&    23.9$\pm$     0.3&  2.9$\pm$  1.7&  0.013$\pm$  0.003\\
TrES-3    b& 0.92& 0.82& 5650& 0.0228&  1.306& 0.170$^a$& 1.90& 1.31& 1638&   1197$\pm$    112&   1082$\pm$     66&    29.2$\pm$     0.2&  1.0$\pm$  0.9&  0.012$\pm$  0.003\\
TrES-4    b& 1.34& 1.83& 6200& 0.0502&  3.554& 0.015$^a$& 0.90& 1.74& 1805&   1770$\pm$    324&   5942$\pm$    989&    25.1$\pm$     0.5&  3.1$\pm$  0.6&  0.018$\pm$  0.005\\
TrES-5    b& 0.89& 0.87& 5171& 0.0245&  1.482& 0.000& 1.78& 1.21& 1484&    804$\pm$     91&    661$\pm$     46&    --- &  5.7$\pm$  1.7&  0.019$\pm$  0.005\\
WASP-01   b& 1.26& 1.47& 6213& 0.0392&  2.520& 0.008$^a$& 0.98& 1.49& 1830&   1868$\pm$    243&   4250$\pm$    551&    25.1$\pm$     0.4&  3.0$\pm$  0.6&  0.019$\pm$  0.005\\
WASP-02   b& 0.85& 0.82& 5170& 0.0309&  2.152& 0.005$^a$& 0.88& 1.06& 1286&    454$\pm$     58&    583$\pm$     55&    24.4$\pm$     0.4&  5.3$\pm$  3.6&  0.016$\pm$  0.004\\
WASP-03   b& 1.11& 1.30& 6340& 0.0305&  1.847& 0.007$^a$& 1.77& 1.35& 2020&   2627$\pm$    501&   2689$\pm$    418&    25.5$\pm$     0.5&  5.8$\pm$  1.6&  0.018$\pm$  0.005\\
WASP-04   b& 0.93& 0.91& 5540& 0.0232&  1.338& 0.003$^a$& 1.25& 1.36& 1673&   1303$\pm$    155&   1941$\pm$    160&    25.8$\pm$     0.3&  4.3$\pm$  2.5&  0.015$\pm$  0.004\\
WASP-05   b& 1.03& 1.09& 5770& 0.0274&  1.628& 0.000& 1.60& 1.17& 1753&   1569$\pm$    236&   1358$\pm$    171&    --- &  4.9$\pm$  1.4&  0.017$\pm$  0.004\\
WASP-06   b& 0.88& 0.87& 5375& 0.0421&  3.361& 0.054& 0.50& 1.22& 1194&    320$\pm$     46&    953$\pm$    115&    25.6$\pm$     0.4&  4.4$\pm$  2.6&  0.012$\pm$  0.003\\
WASP-07   b& 1.32& 1.48& 6520& 0.0624&  4.955& 0.034$^a$& 0.98& 1.37& 1530&    910$\pm$    179&   1753$\pm$    472&    24.4$\pm$     0.6&  1.8$\pm$  0.5&  0.015$\pm$  0.004\\
WASP-08   b& 1.03& 0.94& 5600& 0.0801&  8.159& 0.304$^a$& 2.25& 1.04&  --&    122$\pm$     24&     58$\pm$      2&    24.6$\pm$     0.2&  1.6$\pm$  1.6&  0.019$\pm$  0.005\\
WASP-11   b& 0.83& 0.79& 4980& 0.0435&  3.722& 0.000& 0.46& 1.00& 1020&    182$\pm$     23&    399$\pm$     49&    --- &  5.5$\pm$  2.8&  0.018$\pm$  0.004\\
WASP-12   b& 1.36& 1.60& 6313& 0.0231&  1.091& 0.037$^a$& 1.42& 1.85& 2578&   6864$\pm$   1137&  16544$\pm$   1503&    29.0$\pm$     0.5&  2.1$\pm$  0.4&  0.020$\pm$  0.005\\
WASP-13   b& 1.22& 1.66& 6025& 0.0557&  4.353& 0.000& 0.51& 1.53& 1531&   1047$\pm$    182&   4774$\pm$   1084&    --- &  4.3$\pm$  1.3&  0.017$\pm$  0.004\\
WASP-14   b& 1.35& 1.67& 6462& 0.0372&  2.243& 0.082$^a$& 7.90& 1.63& 2090&   3139$\pm$    697&   1059$\pm$    180&    27.6$\pm$     0.6&  2.1$\pm$  0.6&  0.015$\pm$  0.004\\
WASP-15   b& 1.30& 1.52& 6573& 0.0516&  3.752& 0.038$^a$& 0.59& 1.41& 1676&   1455$\pm$    183&   4873$\pm$    474&    25.3$\pm$     0.3&  2.4$\pm$  0.4&  0.017$\pm$  0.004\\
WASP-16   b& 0.98& 1.09& 5630& 0.0415&  3.119& 0.015$^a$& 0.83& 1.22& 1389&    618$\pm$    101&   1103$\pm$    122&    24.6$\pm$     0.4&  7.0$\pm$  2.0&  0.017$\pm$  0.004\\
WASP-17   b& 1.29& 1.58& 6550& 0.0514&  3.735& 0.039$^a$& 0.48& 1.93& 1755&   1570$\pm$    240&  12289$\pm$   1524&    26.0$\pm$     0.4&  2.3$\pm$  0.5&  0.011$\pm$  0.003\\
WASP-18   b& 1.29& 1.25& 6400& 0.0205&  0.941& 0.007$^a$&10.52& 1.20& 2411&   5617$\pm$    665&    774$\pm$     59&    27.0$\pm$     0.3&  0.5$\pm$  0.5&  0.017$\pm$  0.004\\
WASP-19   b& 0.94& 1.02& 5460& 0.0163&  0.789& 0.002$^a$& 1.14& 1.41& 2077&   3097$\pm$    386&   5405$\pm$    331&    26.9$\pm$     0.3&  8.1$\pm$  2.1&  0.018$\pm$  0.005\\
WASP-22   b& 1.11& 1.22& 6020& 0.0470&  3.533& 0.011$^a$& 0.59& 1.16& 1466&    793$\pm$    106&   1810$\pm$    243&    23.9$\pm$     0.4&  4.3$\pm$  0.5&  0.016$\pm$  0.004\\
WASP-23   b& 0.84& 0.82& 5046& 0.0380&  2.944& 0.000& 0.92& 1.07& 1152&    270$\pm$     53&    336$\pm$     42&    --- &  5.7$\pm$  4.0&  0.016$\pm$  0.004\\
WASP-24   b& 1.15& 1.35& 6297& 0.0362&  2.341& 0.003$^a$& 1.09& 1.38& 1781&   1975$\pm$    195&   3464$\pm$    274&    26.3$\pm$     0.2&  4.7$\pm$  0.4&  0.017$\pm$  0.004\\
WASP-25   b& 1.00& 0.92& 5736& 0.0473&  3.765& 0.000& 0.58& 1.22& 1212&    367$\pm$     47&    943$\pm$    157&    --- &  1.8$\pm$  1.0&  0.016$\pm$  0.004\\
WASP-26   b& 1.10& 1.29& 6034& 0.0398&  2.767& 0.003& 1.03& 1.27& 1623&   1250$\pm$    141&   1957$\pm$    253&    23.6$\pm$     0.4&  6.1$\pm$  0.7&  0.018$\pm$  0.005\\
WASP-30   b& 1.25& 1.39& 6190& 0.0553&  4.157& 0.000&62.50& 0.95& 1474&    830$\pm$     80&     12$\pm$      0&    --- &  2.5$\pm$  0.3&  0.017$\pm$  0.004\\
WASP-31   b& 1.16& 1.25& 6175& 0.0466&  3.406& 0.000& 0.48& 1.55& 1575&    942$\pm$    106&   4731$\pm$    592&    --- &  2.6$\pm$  0.3&  0.012$\pm$  0.003\\
WASP-32   b& 1.10& 1.11& 6100& 0.0394&  2.719& 0.018& 3.60& 1.18& 1560&    986$\pm$    168&    381$\pm$     52&    25.1$\pm$     0.4&  2.5$\pm$  0.6&  0.013$\pm$  0.003\\
WASP-33   b& 1.56& 1.51& 7430& 0.0259&  1.220& 0.000& 3.27& 1.68& 2710&   9288$\pm$    877&   8016$\pm$   1963&    --- &  0.3$\pm$  0.1&  0.017$\pm$  0.004\\
WASP-34   b& 1.01& 0.93& 5704& 0.0524&  4.318& 0.011$^a$& 0.59& 1.22& 1250&    299$\pm$     87&    755$\pm$    148&    23.5$\pm$     0.6&  1.8$\pm$  2.1&  0.017$\pm$  0.004\\
WASP-35   b& 1.07& 1.09& 6072& 0.0432&  3.162& 0.000& 0.72& 1.32& 1450&    778$\pm$     86&   1882$\pm$    299&    --- &  3.2$\pm$  0.7&  0.014$\pm$  0.004\\
WASP-36   b& 1.04& 0.95& 5928& 0.0264&  1.537& 0.000& 2.30& 1.28& 1724&   1435$\pm$    139&   1022$\pm$     76&    --- &  1.1$\pm$  0.8&  0.015$\pm$  0.004\\
WASP-37   b& 0.93& 1.00& 5940& 0.0446&  3.577& 0.000& 1.80& 1.00& 1323&    565$\pm$    128&    315$\pm$     63&    --- &  5.1$\pm$  3.6&  0.009$\pm$  0.002\\
WASP-38   b& 1.20& 1.33& 6150& 0.0752&  6.872& 0.033$^a$& 2.69& 1.09& 1292&    402$\pm$     46&    178$\pm$     13&    23.0$\pm$     0.3&  2.6$\pm$  0.4&  0.013$\pm$  0.003\\
WASP-41   b& 0.94& 0.91& 5546& 0.0403&  3.052& 0.000& 0.92& 1.21& 1235&    433$\pm$     68&    689$\pm$    132&    --- &  4.1$\pm$  1.4&  0.016$\pm$  0.004\\
WASP-42   b& 0.88& 0.86& 5315& 0.0548&  4.982& 0.060& 0.50& 1.08&  995&    177$\pm$     38&    414$\pm$     72&    24.4$\pm$     0.6&  6.9$\pm$  5.7&  0.021$\pm$  0.005\\
WASP-44   b& 0.92& 0.87& 5400& 0.0344&  2.424& 0.000& 0.87& 1.00& 1304&    481$\pm$    127&    556$\pm$     93&    ---            &  3.6$\pm$  4.2&  0.016$\pm$  0.004\\
WASP-45   b& 0.91& 0.94& 5100& 0.0405&  3.126& 0.000& 1.01& 1.16& 1198&    330$\pm$    127&    441$\pm$    236&    ---            &  9.8$\pm$  3.8&  0.023$\pm$  0.006\\
WASP-46   b& 0.96& 0.92& 5600& 0.0245&  1.430& 0.000& 2.10& 1.31& 1654&   1238$\pm$    236&   1011$\pm$    113&    ---            &  2.5$\pm$  1.1&  0.010$\pm$  0.002\\
WASP-47   b& 1.08& 1.15& 5576& 0.0520&  4.159& 0.000& 1.14& 1.15& 1220&    424$\pm$     52&    492$\pm$     55&    ---            &  5.6$\pm$  1.0&  0.023$\pm$  0.006\\
WASP-48   b& 1.19& 1.75& 6000& 0.0344&  2.144& 0.000& 0.98& 1.67& 2030&   3004$\pm$    684&   8549$\pm$   1808&    ---            &  4.0$\pm$  0.6&  0.013$\pm$  0.003\\
WASP-50   b& 0.86& 0.86& 5518& 0.0291&  1.955& 0.000& 1.44& 1.14& 1410&    717$\pm$     85&    646$\pm$     60&    ---            &  6.7$\pm$  4.1&  0.018$\pm$  0.004\\
WASP-52   b& 0.87& 0.79& 5000& 0.0272&  1.750& 0.000& 0.46& 1.27& 1315&    473$\pm$     72&   1659$\pm$    150&    ---            &  2.4$\pm$  2.1&  0.016$\pm$  0.004\\
WASP-54   b& 1.21& 1.83& 6296& 0.0499&  3.694& 0.067& 0.64& 1.65& 1759&   1895$\pm$    270&   8143$\pm$   1206&    26.2$\pm$     0.4&  4.1$\pm$  0.4&  0.015$\pm$  0.004\\
WASP-55   b& 1.01& 1.06& 6070& 0.0533&  4.466& 0.000& 0.57& 1.30& 1290&    482$\pm$     56&   1429$\pm$    210&    --- &  5.2$\pm$  1.3&  0.017$\pm$  0.004\\
WASP-56   b& 1.02& 1.11& 5600& 0.0546&  4.167& 0.000& 0.57& 1.09& 1216&    366$\pm$     48&    765$\pm$     94&    --- &  6.2$\pm$  0.8&  0.018$\pm$  0.004\\
WASP-57   b& 0.95& 0.84& 5600& 0.0386&  2.839& 0.000& 0.67& 0.92& 1251&    414$\pm$    107&    517$\pm$     56&    --- &    ---    &  0.011$\pm$  0.003\\
WASP-58   b& 0.94& 1.17& 5800& 0.0561&  5.017& 0.000& 0.89& 1.37& 1270&    441$\pm$    175&    931$\pm$    345&    --- &  6.9$\pm$  2.8&  0.009$\pm$  0.002\\
WASP-60   b& 1.08& 1.14& 5900& 0.0531&  4.305& 0.000& 0.51& 0.86& 1320&    501$\pm$    159&    721$\pm$    248&    --- &  3.8$\pm$  0.8&  0.015$\pm$  0.004\\
WASP-61   b& 1.22& 1.36& 6250& 0.0514&  3.856& 0.000& 2.06& 1.24& 1565&    959$\pm$    167&    715$\pm$     93&    --- &  2.5$\pm$  0.7&  0.014$\pm$  0.003\\
WASP-62   b& 1.25& 1.28& 6230& 0.0567&  4.412& 0.000& 0.57& 1.39& 1440&    689$\pm$    106&   2336$\pm$    365&    --- &  1.5$\pm$  0.5&  0.016$\pm$  0.004\\
WASP-64   b& 1.00& 1.06& 5550& 0.0265&  1.573& 0.000& 1.27& 1.27& 1689&   1359$\pm$    235&   1728$\pm$    198&    --- &  4.6$\pm$  0.9&  0.014$\pm$  0.003\\
WASP-65   b& 0.93& 1.01& 5600& 0.0334&  2.311& 0.000& 1.55& 1.11& 1480&    807$\pm$    214&    644$\pm$    134&    --- &  6.6$\pm$  4.9&  0.014$\pm$  0.004\\
WASP-66   b& 1.30& 1.75& 6600& 0.0546&  4.086& 0.000& 2.32& 1.39& 1790&   1750$\pm$    396&   1457$\pm$    270&    --- &  2.5$\pm$  0.4&  0.011$\pm$  0.003\\
WASP-67   b& 0.87& 0.87& 5417& 0.0517&  4.614& 0.000& 0.42& 1.40& 1040&    218$\pm$     40&   1021$\pm$    534&    --- &  7.2$\pm$  3.0&  0.019$\pm$  0.005\\
WASP-68   b& 1.24& 1.69& 5910& 0.0621&  5.084& 0.000& 0.95& 1.24& 1490&    812$\pm$    150&   1314$\pm$    253&    --- &  4.4$\pm$  0.4&  0.020$\pm$  0.005\\
WASP-70   b& 1.11& 1.22& 5700& 0.0485&  3.713& 0.000& 0.59& 1.16& 1387&    594$\pm$    111&   1364$\pm$    221&    --- &  4.1$\pm$  0.8&  0.015$\pm$  0.004\\
WASP-71   b& 1.56& 2.26& 6180& 0.0462&  2.904& 0.000& 2.24& 1.46& 2049&   3135$\pm$    670&   2980$\pm$    637&    --- &  2.4$\pm$  0.2&  0.023$\pm$  0.006\\
WASP-72   b& 1.39& 1.98& 6250& 0.0371&  2.217& 0.000& 1.46& 1.27& 2210&   3906$\pm$   1302&   4312$\pm$   1532&    --- &  2.4$\pm$  0.3&  0.014$\pm$  0.004\\
WASP-73   b& 1.34& 2.07& 6030& 0.0551&  4.087& 0.000& 1.88& 1.16& 1790&   1674$\pm$    477&   1198$\pm$    292&    --- &  3.1$\pm$  0.4&  0.018$\pm$  0.005\\
WASP-75   b& 1.14& 1.26& 6100& 0.0375&  2.484& 0.000& 1.07& 1.27& 1710&   1403$\pm$    233&   2115$\pm$    258&    --- &  4.0$\pm$  1.1&  0.017$\pm$  0.004\\
WASP-76   b& 1.46& 1.73& 6250& 0.0330&  1.810& 0.000& 0.92& 1.83& 2160&   3765$\pm$    529&  13705$\pm$   1345&    --- &  1.4$\pm$  0.3&  0.020$\pm$  0.005\\
WASP-77   b& 1.00& 0.95& 5605& 0.0240&  1.360& 0.000& 1.76& 1.21&   --&   1403$\pm$    127&   1167$\pm$     78&    --- &  2.9$\pm$  1.5&  0.017$\pm$  0.004\\
WASP-78   b& 1.33& 2.20& 6291& 0.0362&  2.175& 0.000& 0.89& 1.70& 2350&   5193$\pm$   1030&  16865$\pm$   3698&    --- &  2.8$\pm$  0.5&  0.014$\pm$  0.004\\
WASP-79   b& 1.52& 1.91& 6600& 0.0535&  3.662& 0.000& 0.90& 2.09& 1900&   2171$\pm$    401&  10538$\pm$   2348&    --- &  1.3$\pm$  0.2&  0.016$\pm$  0.004\\
WASP-82   b& 1.63& 2.18& 6500& 0.0447&  2.706& 0.000& 1.24& 1.67& 2190&   3811$\pm$    586&   8573$\pm$    995&    --- &  1.3$\pm$  0.1&  0.018$\pm$  0.004\\
WASP-84   b& 0.84& 0.75& 5300& 0.0771&  8.523& 0.000& 0.69& 0.94&  797&     66$\pm$      9&     85$\pm$      7&    --- &  1.2$\pm$  2.9&  0.015$\pm$  0.004\\
WASP-88   b& 1.45& 2.08& 6430& 0.0643&  4.954& 0.000& 0.56& 1.70& 1772&   1605$\pm$    347&   8285$\pm$   2450&    --- &  1.9$\pm$  0.2&  0.014$\pm$  0.003\\
WASP-90   b& 1.55& 1.98& 6440& 0.0562&  3.916& 0.000& 0.63& 1.63& 1840&   1916$\pm$    410&   8084$\pm$   1790&    --- &  1.4$\pm$  0.2&  0.017$\pm$  0.004\\
WASP-95   b& 1.11& 1.13& 5830& 0.0342&  2.185& 0.000& 1.13& 1.21& 1570&   1134$\pm$    324&   1470$\pm$    275&    --- &  3.2$\pm$  1.7&  0.018$\pm$  0.005\\
WASP-96   b& 1.06& 1.05& 5500& 0.0453&  3.426& 0.000& 0.48& 1.20& 1285&    441$\pm$    115&   1324$\pm$    215&    --- &  3.3$\pm$  2.1&  0.018$\pm$  0.005\\
WASP-97   b& 1.12& 1.06& 5670& 0.0330&  2.073& 0.000& 1.32& 1.13& 1555&    955$\pm$    178&    924$\pm$    133&    --- &  1.7$\pm$  1.0&  0.020$\pm$  0.005\\
WASP-99   b& 1.48& 1.76& 6150& 0.0717&  5.753& 0.000& 2.78& 1.10& 1480&    773$\pm$    181&    336$\pm$     64&    --- &  1.3$\pm$  0.3&  0.020$\pm$  0.005\\
WASP-100  b& 1.57& 2.00& 6900& 0.0457&  2.849& 0.000& 2.03& 1.69& 2190&   3897$\pm$   1611&   5483$\pm$   2205&    --- &  1.1$\pm$  0.2&  0.015$\pm$  0.004\\
WASP-101  b& 1.34& 1.29& 6380& 0.0506&  3.586& 0.000& 0.50& 1.41& 1560&    966$\pm$    167&   3844$\pm$    580&    --- &    ---    &  0.019$\pm$  0.005\\
WTS-1     b& 1.20& 1.15& 6250& 0.0470&  3.352& 0.000& 4.01& 1.49& 1500&    820$\pm$    282&    454$\pm$    137&    --- &  0.8$\pm$  0.9&  0.011$\pm$  0.003\\
XO-1      b& 1.04& 0.94& 5750& 0.0494&  3.942& 0.000& 0.92& 1.21& 1210&    356$\pm$     42&    560$\pm$     84&    --- &  1.0$\pm$  1.3&  0.016$\pm$  0.004\\
XO-2      b& 0.97& 0.99& 5340& 0.0362&  2.616& 0.028$^a$& 0.57& 0.99& 1328&    547$\pm$     66&    953$\pm$     65&    25.2$\pm$     0.3&  8.0$\pm$  1.8&  0.026$\pm$  0.006\\
XO-3      b& 1.21& 1.41& 6429& 0.0453&  3.192& 0.283$^a$&11.83& 1.25& 1729&   1484$\pm$    235&    195$\pm$     21&    27.4$\pm$     0.4&  3.0$\pm$  0.6&  0.012$\pm$  0.003\\
XO-4      b& 1.28& 1.53& 6397& 0.0547&  4.125& 0.002$^a$& 1.55& 1.29& 1630&   1176$\pm$    714&   1259$\pm$    842&    22.3$\pm$     1.9&  2.5$\pm$  0.8&  0.015$\pm$  0.004\\
XO-5      b& 0.91& 1.07& 5370& 0.0494&  4.188& 0.013$^a$& 1.08& 1.09& 1203&    347$\pm$     75&    379$\pm$     76&    23.5$\pm$     0.7&  9.6$\pm$  3.2&  0.016$\pm$  0.004\\
\hline
\end{longtable}
$^a$ The eccentricities are taken from Knutson et al. (2014).

\twocolumn

\label{lastpage}																								     
\end{document}